\documentclass[12pt]{article}
\pdfoutput=1
\usepackage[
top = 2.5cm, 
bottom = 2.5cm, 
left = 2.5cm, 
right = 2.5cm]{geometry}

\usepackage{jheppub}
\usepackage[sort&compress]{natbib}
\usepackage[utf8]{inputenc}
\usepackage{ifpdf}
\usepackage{braket}
\usepackage{graphicx}
\usepackage{caption}
\usepackage{subcaption}
\usepackage{slashed}
\usepackage{feynmf}
\usepackage{epstopdf}
\usepackage{amsmath, amssymb,url,bm,textgreek,upgreek}
\usepackage[svgnames]{xcolor}
\usepackage{hyperref}
\usepackage{dsfont}
\usepackage{float}

\usepackage{tikz} 
\usepackage{tikz-feynman}
\usetikzlibrary{shapes,arrows,positioning,automata,backgrounds,calc,er,patterns}
\usetikzlibrary{decorations.pathmorphing}
\tikzfeynmanset{compat=1.0.0}

\hypersetup{linkcolor=DarkBlue,citecolor=magenta}
\pdfoutput=1
\makeatletter
\def\@fpheader{\relax}
\makeatother

\def\be{\begin{eqnarray}}
\def\ee{\end{eqnarray}}

\def\tr{\operatorname{tr}}

\def\Vol{\operatorname{Vol}}

\def\tr{\operatorname{tr}}

\def\d{{\rm d}}

\def\sgn{\operatorname{sgn}}

\def\sD{\slashed{\rm D}}

\def\id{\mathds{1}}
\def \bea {\begin{eqnarray}}
\def \eea {\end{eqnarray}}
\def \nn {\nonumber}

\newcommand{\oR}{\mathring{R}}

\def \del {\partial}

\def \cN {\mathcal{N}}
\def \t {\widetilde}

\title{Superconformal Models for Graphene \\
and Boundary Central Charges}
\author{Christopher P. Herzog$^{a,b}$,}
\author{Kuo-Wei Huang$^{b,c
}$,}
\author{Itamar Shamir$^a$,}
\author{and Julio Virrueta$^b$}

\affiliation{
$^a$ Mathematics Department, King's College London, \\
The Strand, London,  WC2R 2LS, UK \\
$^b$ C. N. Yang Institute for Theoretical Physics, Department of Physics and Astronomy,\\
Stony Brook University, Stony Brook, NY 11794, USA \\
$^c$ Perimeter Institute for Theoretical Physics, Waterloo, Ontario N2L 2Y5, Canada
}

\abstract{
In the context of boundary conformal field theory, 
we investigate whether the boundary trace anomaly can 
depend on marginal directions in the presence of supersymmetry.  
Recently, it was found that a graphene-like non-supersymmetric conformal field theory with 
a four-dimensional bulk photon and a three-dimensional boundary electron
has two boundary central charges that depend on an exactly marginal direction, namely the gauge coupling. 
In this work, we supersymmetrize this theory, paying special attention to the boundary terms required by supersymmetry. 
We study models with 4, 8, and 16 Poincar\'e supercharges in the bulk, half of which are broken by the boundary.
In all cases, we find that at all orders in  perturbation theory, the gauge coupling is not renormalized, providing strong evidence
that these
theories are boundary conformal field theories. 
Moreover, the boundary central charges depend on the coupling. 
One possible exception to this dependence on marginal directions is that the
 difference between the two charges is coupling independent at one-loop 
in the maximally supersymmetric case. 
In our analysis, a possible boundary Chern-Simons term is incorporated by a bulk $\theta$-term. 
}

\preprint{YITP-18-11}

\begin{document}
\maketitle
\setcounter{page}{2}

\newpage
\section{Introduction}

This research is motivated by a desire to understand the structure of quantum field theory.  
Our working hypothesis is that much new can be learned by focusing on quantum field theory in the presence of boundaries and defects.  
Indeed, there has been enormous progress 
associated with boundary quantum field theory, gravitational systems with a boundary, and boundary effects in string theory. 
D-branes, i.e. the boundaries of fundamental strings, helped lead to the second superstring revolution in the late 90s by providing non-perturbative insight into the various string theories.  
In gauge-gravity duality, a central role is played by the conformal boundary of anti-de Sitter space in a gravitational theory.  
Entanglement entropy, which has helped refine our notion of renormalization group flow in quantum field theory while at the same time providing insight into black hole physics, is often defined spatially, with a central role played by the entangling surface that separates two regions. 
Boundary effects are also essential for understanding condensed matter systems such as topological insulators.

As fixed points of the renormalization group flow, conformal field theories are important landmarks in the space of quantum field theory more generally.  While the stress tensor of a CFT is traceless classically, on a curved space-time there are anomalies that provide important ways of characterizing CFTs and renormalization group flows between them.  In four space-time dimensions, there are two such anomaly coefficients, often called $a$ and $c$.  
In our convention, the central charge $a$ multiplies the Euler density, while $c$ multiplies the square of the Weyl curvature.  
Both numbers can be used to check conjectured dualities between different quantum field theories.  
Remarkably, $a$-charge orders QFTs along renormalization group flows \cite{Komargodski:2011vj}, with $a_{\rm UV} > a_{\rm IR}$.  On the other hand, the $c$-charge determines the coefficient of the stress tensor two-point function in general 4d CFTs
\cite{Osborn:1993cr}. 

In the presence of a 2+1 dimensional boundary, two additional anomaly coefficients appear, which we shall call $b_1$ and $b_2$.  
The complete classification based on the Wess-Zumino consistency condition was given recently in \cite{Herzog:2015ioa} and the trace of the stress tensor takes the following general form:  
 \be
\label{anomalies}
 \langle {T^\mu}_\mu \rangle^{(4{\rm d})} 
= \frac{1}{16 \pi^2} (c W_{\mu\nu \lambda \rho}^2 - a E_4 ) + \frac{\delta(x^n)}{16 \pi^2} (a E_4^{\rm (bry)} - b_1 \tr \hat K^3 - b_2 h^{\alpha\beta} \hat K^{\gamma \delta} W_{\alpha  \gamma \beta \delta} ) \ ,
 \ee
 where $E_4$ is the Euler density and $W_{\mu\nu\lambda \rho}$ the Weyl curvature; $\delta(x^n)$ is a Dirac delta function with support on the boundary. 
We have ignored the total-derivative anomaly in \eqref{anomalies}, $\Box R$, which is scheme-dependent.
Note that the Euler density has a boundary contribution $E_4^{\rm (bry)}$.  
We refer the reader to \cite{Herzog:2015ioa} for detailed discussions related to the Euler boundary term $E_4^{\rm (bry)}$, which has a delicate connection to the universal  entanglement entropy across a sphere.\footnote{%
See also ref.\ \cite{Herzog:2016kno} which reproduces the universal entanglement entropy with a more general shape of entangling surface via a dimensional reduction of the boundary conformal anomaly.
} 
The general property $a_{\rm UV} > a_{\rm IR}$ should not be violated when a boundary is present.
To describe the boundary contributions,  we construct a projector onto the boundary metric $h_{\mu\nu} = g_{\mu\nu} - n_\mu n_\nu$, with $n_\mu$ a unit, outward normal vector to the boundary.  Then $\hat K_{\mu\nu} = K_{\mu\nu} - \frac{K}{3} h_{\mu\nu}$ is the traceless part of the extrinsic curvature.  
The $b_1$- and $b_2$-anomalies will be the main focus of the present paper.

Given the importance of $a$ and $c$, an effort should be made to understand constraints on and properties of the two new coefficients $b_1$ and $b_2$.  A certain amount is known already.  
The coefficients are proportional to two- and three-point functions of the displacement operator 
\cite{Herzog:2017kkj,Herzog:2017xha}, i.e.\ the operator conjugate to the position of the boundary.  
The coefficients have been computed for free theories \cite{Fursaev:2015wpa, Fursaev:2016inw, JM, Moss} and perturbatively for one interacting theory \cite{Herzog:2017xha}.  By reflection positivity of the displacement two-point function, one has the inequality $b_2 \geq 0$.  
In free theories, one has universally that $b_2=8c$ while such a relation can be violated by introducing boundary interactions \cite{Herzog:2017xha}. 

One of the most interesting stories about these anomaly coefficients concerns their dependence on marginal couplings.  Certain special CFTs in $d$ space-time dimensions belong to larger families
parametrized by a set of marginal couplings.  (Marginal means they source operators with 
a scaling dimension $\Delta = d$ that is independent of the coupling strength.)
Wess-Zumino consistency implies that $a$ is independent of these couplings \cite{Osborn1991weyl}.  The situation for $c$ is murkier.  On the one hand, in the presence of supersymmetry, a particular linear combination of $a$ and $c$ is fixed by an anomaly in the $R$-symmetry current, which also must be independent of these couplings \cite{Anselmi:1997am}. Thus, for a supersymmetric theory, $c$ must be independent.  On the other hand in 4d without a boundary, no example of a non-supersymmetric CFT with marginal directions is known.  The logical possibility remains that if one found a non-supersymmetric family of CFTs with a marginal coupling, $c$ could depend on that 
coupling.\footnote{%
 See refs.\ \cite{Nakayama:2017oye,Bashmakov:2017rko} for recent discussions of this issue.
 }

The situation with a boundary is richer.  It turns out there is a remarkably simple non-supersymmetric CFT with a boundary and an exactly marginal coupling. 
  The theory contains a 4d photon and a 3d electron, and as such is a close cousin of graphene.  
   (For earlier work on this theory, see refs.\
 \cite{Herzog:2017xha,Gorbar:2001qt,SJReystrings,Kaplan:2009kr,Kotikov:2013eha,Kotikov:2016yrn,Hsiao:2017lch}.) 
  The gauge coupling, or equivalently the charge of the electron, is marginal.  An essential difference between this boundary CFT and field theoretic models of graphene (see e.g.\ ref.\ \cite{Son:2007ja}) is that in our theory, the electron and photon travel with the same relativistic dispersion relation, while in real world graphene, the electron travels about 300 times slower. That said, the speed of the electron in graphene has a beta function; our theory could be thought of as the ultimate IR fixed point of real world graphene, albeit a fixed point one is far from being able to realize in the lab. 
  From a field theoretic standpoint, this fixed point theory may nevertheless be a useful and tractable starting point for approximating real world graphene \cite{Kotikov:2013eha,Kotikov:2016yrn,Hsiao:2017lch}.
(In the context of graphene, the fact that the charge of the electron has a vanishing beta function is discussed in various reviews, see e.g.\ ref.\ \cite{Vozmediano:2010fz}.) 

In this graphene-like theory, the boundary anomaly coefficients are more interesting than the bulk ones.
As the interactions are confined to the boundary, $a$ and $c$ are fixed by their values for a free photon.  The boundary coefficients $b_1$ and $b_2$ however can be shown to depend perturbatively on the charge of the electron \cite{Herzog:2017kkj,Herzog:2017xha}.

A natural question is whether supersymmetry can further constrain the coefficients $b_1$ and $b_2$ like it does for $c$.  While ultimately one should find a general argument based on the multiplet structure of supersymmetric theories with boundary, in the style of refs.\ \cite{Belyaev:2008xk,Drukker:2017dgn,Drukker:2017xrb,DiPietro:2015zia}, 
a simpler approach is to study a couple of examples, to see what types of behaviors are possible.   In this paper, we consider supersymmetric versions of graphene 
with four ($\cN=1$), eight ($\cN=2$), and sixteen supercharges ($\cN=4$) in the bulk. 
The presence of the boundary breaks half of the supersymmetries. In each case, we consider a free abelian gauge multiplet which is coupled to matter multiplets localized on the boundary. The matter fields form multiplets of the effective supersymmetry on the boundary which is 3d $\cN=1$, 2 and 4 respectively. We construct explicitly the theories with $\cN=1$ and $\cN=2$ in the bulk, emphasizing the role of boundary terms necessary for off-shell supersymmetry, and obtaining supersymmetric boundary conditions. Building on this, we obtain results also for $\cN=4$ in the bulk.

We consider in detail the effect of the $\theta F \wedge F$ term. In the presence of a boundary, the symmetry for shifting $\theta$ by $2\pi$ is lost and the boundary Chern-Simons term is essentially the integer part of $\theta/2\pi$. Normally, to couple the gauge field to charged fields on the boundary one chooses a Neumann boundary condition $F_{n A}=0$ (where $A$ is an index tangent to the boundary) which keeps the effective boundary gauge field unconstrained. Introducing a $\theta$-term produces a Robin type boundary condition $F_{n A} +\tan (\alpha) \t F_{n A}=0$, where $\t F_{n A}$ is the dual field strength $\frac{1}{2} \epsilon_{n ABC}F^{BC}$, and $\tan (\alpha) = \frac{\theta g^2}{4\pi^2}$. This change in the boundary condition has the effect of screening the gauge coupling $g \to g \cos(\alpha)$.\footnote{%
See ref.\ \cite{Dudal:2018mms,Mulligan:2013he,Marino:1992xi} for related work.
}

As mentioned above, the anomaly coefficients $b_1$ and $b_2$ are obtained from the two- and three-point functions of the displacement operator, which in turn is obtained as the boundary value of the stress tensor component $T_{nn}$. Noting that this component depends only on the bulk fields, the leading correction to the free theory result comes from the one-loop correction to the propagators of the bulk fields. As a consequence of supersymmetry, the corrections to the propagators are specified by a single coefficient, as we demonstrate by explicit computation.

We find that in all three examples, the gauge coupling continues to be exactly marginal.  Morever, both $b_1$ and $b_2$ depend perturbatively on this coupling.  
Thus, the conclusion is that the  situation for $b_1$ and $b_2$ is rather different than the situation for $c$.  
While $c$ is constrained by supersymmetry to be independent of marginal couplings, $b_1$ and $b_2$ are not.
Our results provide a counter-example to any general argument that $b_1$ or $b_2$ must be independent of marginal couplings in the presence of supersymmetry.  An interesting caveat is that with ${\mathcal N}=4$ supersymmetry, there may be a particular combination, $b_1 - b_2$, of the charges which remains independent of the coupling.\footnote{%  
Of course there could still be special cases where supersymmetrizing a given theory does lead to $b_1$ and $b_2$ which
are independent of marginal couplings.
}

We are making the assumption that our super-graphene theories are examples of boundary conformal field theory, where the full conformal group is broken from $O(4,2)$ to $O(3,2)$ by the presence of a boundary.  The assumption is based on an all orders perturbative argument that the beta function for the gauge coupling vanishes as well as power counting arguments about other possible couplings that could be generated at loop level, but the assumption could be wrong.  There could be non-perturbative corrections to the beta function.  The theory may be unstable with respect to a symmetry breaking phase transition, for example one that spontaneously breaks the $U(N_f)$ flavor symmetry, although one may reasonably hope that for sufficiently small coupling, the theory remains stable.\footnote{%
 The hope is based on a relationship to three-dimensional QED with $N_f$ flavors \cite{Kotikov:2016yrn} where there may be a similar symmetry breaking below a critical $N_f$, with the identification $N_f \sim 1/g$. Note that there is a closer relationship between these graphene-like theories and three dimensional QED than with its four dimensional cousin.  
 In the large number of flavors limit, three dimensional QED is expected to flow to a conformal fixed point where the Feynman rules become very similar to those of our theories.
 }  
These issues about stability and non-perturbative effects deserve further study, but lie outside the scope of the present work.

The structure of this paper is as follows.  
Section \ref{sec:models} discusses the various graphene-like models.
Section \ref{sec:nonsusy} contains a brief review of the non-supersymmetric 
graphene-like model employed in ref.\ \cite{Herzog:2017xha} along with a new discussion of the effect of the $\theta F\wedge F$ term in the action.
In section \ref{sec:N1}, we introduce our theory with four bulk supercharges, dubbed ${\mathcal N}=1$ super graphene.  In section \ref{sec:N2}, we continue with our eight supercharge theory, ${\mathcal N}=2$ super graphene. Section \ref{sec:props} contains the calculation of perturbative corrections to the coefficients $b_1$ and $b_2$ along with a detailed discussion of propagators.  
In section \ref{sec:oneloop}, we consider a one-loop analysis of super graphene.  
We show that the theories are perturbatively scale invariant, and also calculate one-loop self-energies of the bulk fields, needed for the calculations in section \ref{sec:props}.

We end with a short discussion containing several potential future projects. 
Appendix A provides details about our conventions for fermions.
Appendix B lists relevant Feynman rules needed for our one-loop computations.

\section{Ultrarelativistic Models of Graphene}
\label{sec:models}

\subsection{Non-Supersymmetric Model}
\label{sec:nonsusy}

The non-supersymmetric model of graphene (mixed dimensional QED) mentioned in the introduction and used in ref.\ \cite{Herzog:2017xha}
has the following action: 
\be
S_{\rm tot}= \int_{\cal M}  \d^4 x \, \left( - \frac{1}{4} F^{\mu\nu} F_{\mu\nu} + \frac{g^2\theta}{16\pi^2} F^{\mu\nu} \t F_{\mu\nu} \right)
+  \int_{\partial {\cal M}} \d^3 x  \left( i\t \psi \slashed{D} \psi  \right) \ .
\ee
The notation requires some unpacking.  
Greek indices $\mu, \nu$ are bulk while Roman indices $A, B$ are reserved for the boundary.
We will denote the index $n$ as the direction normal to the boundary and  the space ${\cal M}$ corresponds to $x^n >0$ while the boundary $\partial {\cal M}$ is the locus $x^n = 0$. 
We raise and lower indices with a Minkowski tensor $\eta^{\mu\nu}$ with mostly plus signature.   
The Maxwell field strength $F_{\mu\nu} = \partial_\mu A_\nu - \partial_\nu A_\mu$ is constructed in the usual way, and we also use the dual field strength $\t F_{\mu\nu} = \frac{1}{2} \epsilon_{\mu\nu\rho\sigma} F^{\rho\sigma}$.  (Note that in ref.\ \cite{Herzog:2017xha}, $\theta$ was set to zero.)
We let $D_\mu = \partial_\mu - i g A_\mu$ and $\sD =\Gamma^A {\rm D}_A$. 
The 4d gamma matrices $\gamma^\mu$ and 3d gamma matrices $\Gamma^A$ satisfy the usual Clifford algebra 
$\{ \gamma^\mu, \gamma^\nu \} = -2\eta^{\mu\nu}$ and $\{\Gamma^A, \Gamma^B \} = -2 \eta^{AB}$, and $\gamma^5 = \gamma^0 \gamma^1 \gamma^2 \gamma^3$. Additionally, $\t \psi=\psi^\dagger \Gamma^0$ is our notation for a 3d barred spinor.  The standard bar notation, $\bar\lambda$, is reserved for 4d spinors.  
More details about our conventions regarding spinors can be found in Appendix \ref{app:spinor}.

Let there be $N_f$ fermions.  In earlier work \cite{Herzog:2017xha,Herzog:2017kkj}, $N_f$ was assumed to be an even number  to avoid generating a parity anomaly and corresponding induced Chern-Simons term on the boundary.  
We relax this constraint here.  
A Chern-Simons term $\frac{k}{4\pi} A \wedge F$ on the boundary integrates to $\frac{k}{4\pi} F \wedge F$ in the bulk and can be absorbed by a shift of $\theta$.
Note that in the presence of a boundary,
the familiar symmetry of shifting $\theta$ by $2\pi$ is lost since it follows from the quantization of $\int F \wedge F$ on a closed manifold. (Indeed, a way to restore the symmetry is to augment the transformation rule by a shift of the boundary Chern-Simons level \cite{Seiberg:2016gmd}.) 

Shifts in Chern-Simons terms are typically generated through loop effects.  While we do not calculate the shifts -- indeed we cannot in our dimensional regularization scheme -- it is on the one hand well known how to do so using other regularization schemes, e.g.\ Pauli-Villars, and on the other not particularly useful given the ability to shift $\theta$ to whatever value we desire.  Our philosophy is to incorporate the possibility of such shifts by a suitably chosen $\theta$-term.  The $\theta$ in our action is thus to be interpreted as one that includes all of the one loop shifts to the Chern-Simons level and that zeros out the quantum corrected Chern-Simons term on the boundary.

A generic variation of the bulk degrees of freedom leads to the boundary term
\be
\label{nonsusybry}
- \delta A_A \left( F^{nA} - \frac{g^2 \theta}{8 \pi^2} \epsilon^{nABC} F_{BC} + g J^A \right)
\ee
where $J^A = \tilde \psi \Gamma^A \psi$ is the boundary charge current.  In order to have boundary interactions between $A_B$ and $\psi$, the variation $\delta A_B$ should be unconstrained.  Vanishing of the boundary term implies instead a Robin type constraint on $F^{nA}$.
Let us define an angle $\alpha$ associated with this mixing by 
\be
\label{tanalphadef}
\tan (\alpha) \equiv \frac{g^2 \theta}{4\pi^2} \ .
\ee
The boundary condition is then written as\footnote{%  
This condition is reminiscent of a similar effect in ref.\ \cite{Seiberg:1999vs, Leigh:1989jq} 
in which a constant $B$-field background for open strings generates interpolating boundary conditions. 
We thank S.~Murthy for discussion on this point.
}
\be
\label{nonsusybrycond}
\cos (\alpha)F_{nA}  -  \sin (\alpha) \t F_{nA} =  -  g \cos (\alpha) J_A \ .
\ee
This form suggests that we can use the $SL(2,\mathbb{R})$ symmetry of free Maxwell theory to define a new potential $A^\theta_\mu$ whose field strength satisfies the $\theta=0$ boundary condition $F^\theta_{nA}=0$. It is interesting to note that the limit $\theta \to \infty$ corresponds to Dirichlet boundary conditions for the gauge field.  Such boundary conditions decouple the boundary degrees of freedom from the gauge interaction. We therefore anticipate that corrections corresponding to boundary interactions vanish in the $\theta \to \infty$ limit. Indeed, in the $A^\theta$ frame the full boundary condition \eqref{nonsusybrycond} takes the form $F^\theta_{nA} = - g \cos (\alpha) J_A$. The effective coupling is hence $g \cos(\alpha)$ which vanishes in the limit $\theta\to\infty$.

In ref.\ \cite{Herzog:2017xha}, through a one-loop computation, the $\beta$-function of this theory was found to 
vanish. 
In fact, through standard Ward identity and non-renormalization arguments, which we will review in the supersymmetric case later and which hold for arbitrary $\theta$, this model is expected to be exactly conformal in 4d \cite{Herzog:2017xha}. 

The bulk central charges for this model do not depend on $\theta$ or the coupling.
The boundary central charges $b_1$ and $b_2$ are
\be
\label{b1mqed}
b_{1(\rm{Mixed~QED})}&=& {8 \over 35} \Big(2-{3g^2 \cos^2 \alpha \over 8} N_f + {\cal O}(g^4) \Big) \ ,
\\
\label{b2mqed}
b_{2(\rm{Mixed~QED})}&=& {2 \over 5} \Big(2-{g^2 \cos^2 \alpha \over 4}N_f + {\cal O}(g^4) \Big) \ , 
\ee 
both of which depend on $g$.  
While only the $b_2$ result was computed in ref.\ \cite{Herzog:2017xha} (and only in the special case $\theta = 0$) using the results from refs.\ \cite{Herzog:2017xha,Herzog:2017kkj} (or details from later sections of this paper), it is straightforward to compute $b_1$ as well. 
(At zeroth order, these charges are determined by the 4d Maxwell theory with a boundary and are independent of the choice of boundary condition, $F^{nA} = 0$ or $F^{AB} = 0$.\footnote{%
The boundary condition $F_{nA}=0$ on the gauge field is sometimes called ``absolute''.  The Dirichlet-like condition $F_{AB}=0$ on the other hand is called ``relative''.
}%
)  
One of the main motivations of the present work is to generalize  
\eqref{b1mqed} and \eqref{b2mqed} to supersymmetric theories. 

\subsection{${\mathcal N}=1$ Super Graphene}
\label{sec:N1}

Before writing down the action for ${\mathcal N}=1$ super graphene, it is useful to make some general remarks about how the presence of a boundary breaks the 4d ${\mathcal N}=1$ SUSY algebra, 
\be
\{ Q , \bar Q \} = 2i \gamma^\mu \partial_\mu \ ,
\ee
down to a 3d ${\mathcal N}=1$ SUSY subalgebra.  
Here $Q$ is a Majorana supercharge, $\bar Q$ is defined by $Q^T C$ and $C$ is the charge conjugation matrix (see Appendix \ref{app:spinor}). Additionally, the algebra has an R-symmetry acting on $Q$ by $e^{\eta\gamma^5} Q$ ($\gamma^5$ is imaginary in our conventions).

The presence of the boundary breaks translation invariance in the normal direction and as a consequence we can preserve at most half of the bulk supersymmetries. We are therefore looking for a subalgebra consisting of two supercharges, which includes only the tangential translations $\partial_A$. We now show the subalgebra is defined by introducing projectors $\Pi_\pm$ such that 
 \be
 \label{Pipmdef}
\Pi_\pm = \frac{1}{2} \left( 1 \pm \beta \right) \ , ~~~\beta\equiv i \gamma^n \gamma^5 e^{\eta \gamma_5}\ ,
\ee 
along with their barred conjugates $\overline \Pi_\pm = C^{-1} \Pi_\pm^T C=\gamma^0 \Pi_\pm^\dagger \gamma^0$.
We choose the 3d subalgebra to be generated by $Q_+$ and $\bar Q_+$ such that $\Pi_+ Q_+ = Q_+$.
As we are dealing with a conformal theory with an unbroken R-symmetry in the bulk, we may use the R-symmetry to set the real parameter $\eta$ to zero. In view of more general applications, e.g.\ two planar boundaries with two independent parameters 
$\eta$ and  $\eta'$, we will keep the $\eta$ parameter in what follows.

To preserve the subalgbra, the projection operators must act on the gamma matrices as
\be
 \label{n1_subalgebra}
\Pi_+ \gamma^\mu \overline\Pi_+ = \delta^\mu_A \Pi_+ \gamma^A \ .
\ee
As a result, 
in addition to the usual suite of projection operator relations, $\Pi_+ + \Pi_-=1$ and $\Pi_+ \Pi_-=0$, 
the projectors also satisfy
\be
\label{commrel}
\overline{\Pi}_\pm \gamma^A = \gamma^A \Pi_\pm \ , \; \; \; \overline{\Pi}_\pm \gamma^n = \gamma^n \Pi_\mp \ , 
\; \; \; \Pi_\pm \gamma^5 = \gamma^5 \Pi_\mp \ .
\ee
From these commutation relations, one can derive the form (\ref{Pipmdef}).

It is noteworthy that the tangential gamma matrices $\gamma^A$ do not commute with the projectors, and hence cannot be identified with their 3d counterparts. It will be useful to find objects which do possess this property. With this goal in mind, let us define $\t\gamma^\mu = e^{-\eta \gamma^5} \gamma^\mu$. The definition can be understood as conjugation with the R-symmetry operator $ e^{\eta \gamma^5/2}$. It is straightforward to check that $\Pi_{\pm} \t \gamma^A = \t\gamma^A \Pi_\pm$ 
and $\Pi_\pm \t\gamma^n = \t\gamma^n \Pi_{\mp}$. 
Any expression containing projectors can then be easily converted to 3d according to the rule 
$ 
\Pi_{\pm} \t \gamma^A = \pm \Gamma^A
$,
and a rule for associating a 3d barred spinor $\t \lambda$ with $\lambda^\dagger \t\gamma^0 = \bar\lambda e^{\eta\gamma^5}$, satisfying $\widetilde {\Pi_\pm \lambda} = \t \lambda \Pi_\pm$. 
From the definition of the barred spinor we can also identify the 3d charge conjugation matrix with $\t C = e^{\eta\gamma^5} C$ by requiring that $\t \lambda = \lambda^T \t C$. The charge conjugation matrix satisfies the relation $\Pi_\pm^T \t C = \t C \Pi_\pm$. 

With these preliminaries, we are ready to write down the action for ${\mathcal N}=1$ supergraphene.  We divide the action into bulk and boundary contributions:
\be
S_{\rm tot} = S_{\rm bulk} + S_{\rm bry} \ .
\ee
The 4d bulk contains a photon described by a vector field $A_\mu$ and its super partner, a photino, described by a Majorana spinor $\lambda$.  
We also introduce a real scalar auxiliary field $D$. The corresponding action is
\be \label{bulk_action}
S_{\rm bulk} =   \int_{\cal M} \d^4 x \left( - \frac{1}{4} F^{\mu\nu} F_{\mu\nu}  + \frac{g^2\theta}{16\pi^2} F^{\mu\nu} \t F_{\mu\nu} + \frac{i}{2} \bar \lambda 
\slashed{\partial}
\lambda + \frac{1}{2} D^2 \right) \ ,
\ee where $\slashed{\partial}=\gamma^\mu \partial_\mu$ and $\t F_{\mu\nu} = \frac{1}{2} \epsilon_{\mu\nu\rho\sigma} F^{\rho\sigma}$. 
The bulk action preserves four supercharges, with the following supersymmetry transformations:
\be \label{bulk_var}
\delta A_\mu &=& - i \bar \epsilon \gamma_\mu \lambda \ , \\
\delta \lambda &=& 
 \left(\frac{1}{2} F_{\mu\nu} \gamma^{\mu\nu}-\gamma^5 D \right) \epsilon \ , \\
\delta D &=& i \bar \epsilon \gamma^5 \slashed{\partial}  \lambda \ , 
\ee
where $\gamma^{\mu\nu} = \frac{1}{2}[\gamma^\mu,\gamma^\nu]$.  Through Noether's theorem, the SUSY generators $Q$ and $\bar Q$ are related to $\epsilon$ and $\bar \epsilon$ in the usual way.
In the presence of a boundary, 
the preserved subalgebra is parametrized by a spinor variable $\epsilon$ satisfying the condition 
$\Pi_+\epsilon = \epsilon$.  

The 3d boundary contains an electron described by a Dirac spinor $\psi$ and a selectron described by a complex scalar field $\phi$, along with a complex auxiliary field $F$.
The interactions between the photon, photino, electron and selectron are constrained to the boundary:
\be
\label{n1bry}
S_{\rm bry} &=& \int_{\partial {\cal M}} \d^3 x \bigg( i\widetilde \psi \sD \psi - |{\rm D}_A \phi|^2
+ |F|^2 + i  g\big(\t \lambda_+ \psi \phi^*  -  \t \psi  \lambda_+ \phi \big) 
\nonumber \\
&& ~~~~~~~~~~~~~   ~~~~~~~~~~~~~~~~~~~~~~~~~~~~~~~ - \frac{1}{4} \bar \lambda \gamma^5  e^{\eta \gamma^5} \lambda
 - \frac{g^2 \theta}{8\pi^2} \t\lambda_+\lambda_+ \bigg) \ ,
\ee
where $\lambda_+ = \Pi_+ \lambda$.

The terms $\bar \lambda \gamma^5  e^{\eta \gamma^5} \lambda$ and $\t\lambda_+\lambda_+$ in the boundary action are necessary for supersymmetry. 
On its own, the action \eqref{bulk_action} is not invariant under the supersymmetry variations \eqref{bulk_var} in the presence of a boundary, even if the variations are restricted to the subalgebra. The problem is that supersymmetric Lagrangians are invariant only up to a total derivative, thus leading to a boundary term. The added terms $\bar \lambda \gamma^5  e^{\eta \gamma^5} \lambda$ and $\t\lambda_+\lambda_+$ offset the variation coming from the bulk \cite{Belyaev:2005rs,Belyaev:2008xk,DiPietro:2015zia}, but only provided we restrict to the subalgebra. The terms are thus a manifestation of our inability to preserve all four supercharges of the bulk in the presence of a boundary.\footnote{%
 The boundary action can be obtained more systematically via superspace methods.  One path is to find a superspace
 extension of the normal coordinate which is invariant under the preserved supersymmetry.  Integrating the $W^\alpha W_\alpha$ superfield over a supersymmetrized chiral Heaviside theta function leads to the  fermion bilinear boundary terms \cite{Drukker:2017xrb}.
 ($W_\alpha$ is the field strength superfield whose lowest component is the photino.)
Similar procedures can be used in the ${\mathcal N}=2$ case we describe below.
} 

The remaining terms in $S_{\rm bry}$ can be motivated by considering the multiplet structure of 3d ${\mathcal N}=1$ SUSY.
The most basic such multiplet is a scalar multiplet, which consists of $\phi$, $\psi$ and $F$. 
As in graphene, the matter fields are electrically charged and interact with the effective Maxwell field on the boundary, namely $A_A$. To extend this interaction in a supersymmetric fashion we argue as follows. Under the subalgebra derived above, the bulk multiplet decomposes into two multiplets of the preserved symmetry. The multiplet of interest to us includes the effective gauge field as well as a fermion $\lambda_+ = \Pi_+ \lambda$, with variations given by 
\be \label{lambda_A_mult}
\delta A_A = - i \, \t\epsilon \,\Gamma_A \lambda_+ \ ,
\qquad
\delta \lambda_+ = \frac{1}{2} \Gamma^{AB} \epsilon F_{AB} \ .
\ee 
Since this multiplet is identical to the regular gauge multiplet of 3d $\mathcal{N}=1$ supersymmetry, 
the 3d boundary action takes a standard form when written in terms of $\lambda_+$ and the boundary value of $A_A$.

The off-shell SUSY transformations of the boundary multiplet are given by 
\be
\delta \phi &=&  -\t \epsilon \psi\ , \\
\delta \psi &=&  i\Gamma^A \epsilon {\rm D}_A \phi- F \epsilon \ , \\
\delta F &=&\t \epsilon  \,\frac{\partial {\mathcal L}}{\partial \t \psi}= 
i\, \t \epsilon \, {\rm \sD} \psi  - i g\, \t \epsilon \,\lambda_+ \phi\ . %
\ee
By off-shell SUSY we mean not only that the SUSY transformations close off-shell but also that neither boundary conditions nor equations of motion need be applied to have an invariant action $S_{\rm tot}$. We treat boundary conditions and equations of motion on an equal footing.

 %%%%%%%%%%%%%%%%%
 %%%%%%%%%%%%%%%%%
\subsubsection*{Boundary Conditions}

Let us first consider a purely classical analysis of the boundary conditions.
Compared with eq.\ (\ref{nonsusybry}), a generic variation of the bulk degrees of freedom leads to an additional boundary 
term, 
\be \label{bry_var}
-\delta A_A \left(F^{nA} - \tan(\alpha)  \t F^{nA}  + g J^A \right)  - \delta\t \lambda_+ \Big(\lambda_-  + \tan(\alpha) \lambda_+ - i g ( \psi \phi^* - \psi^c \phi) \Big) \ ,
\ee
where $\psi^c = - \t C^{-1} \t\psi^T$ is the Majorana conjugate,\footnote{%
Note that $\psi$ is not Majorana since it's complexified so $\psi^c \neq \psi$.
}
$\tan(\alpha)$ is defined in eq.\ (\ref{tanalphadef}), and the dual field strength is given by
$\t F_{\mu\nu} = \frac{1}{2} \epsilon_{\mu\nu\rho\sigma}F^{\rho\sigma}$.
We define 
\be
\lambda_- \equiv \gamma^5 \Pi_- \lambda
\ee
 such that $\Pi_+ \lambda_- = \lambda_-$, and the electric current on the boundary is given by 
\be
J^A = \t \psi \, \Gamma^A \psi  + i \left[({\rm D}^A \phi)^* \phi  -  \phi^* ({\rm D}^A \phi)\right]\ .
\ee

Two physical considerations lead to a natural choice of boundary conditions.  As in the non-supersymmetric case, we want to allow for non-trivial electric interactions between the bulk gauge field and the charged boundary fields. Interactions require an unconstrained $\delta A_B$ on the boundary and suggest we consider the alternate boundary condition in which 
$F_{nA}$ is constrained. 

Second, the boundary conditions that we choose should be consistent with the preserved supersymmetries, 
i.e.\ no further constraints are generated through super-transformations.
We explained above that the pair $(A_A,\lambda_+)$ is a multiplet of the effective 3d $\mathcal{N}=1$ on the boundary. In the same way $\lambda_-$ and $F_{nA}$ are related by supersymmetry. Indeed,
\begin{align}
&\delta A_n = \t\epsilon \,\lambda_- \ ,
\\
&\delta \lambda_- = \epsilon D + i \Gamma^A \epsilon F_{nA} \ ,
\\
&\delta D = - i \, \t\epsilon \, \Gamma^A \partial_A \lambda_- - \t\epsilon \,\partial_n \lambda_+ \ .
\end{align}
The boundary conditions must be consistent with the multiplet structure $(F^{nA}, \lambda_-)$ and $(\t F^{nA}, \lambda_+)$.
With $\theta = 0$, the condition on $F_{nA}$ is related to a condition on the photino, $\lambda_-$.  Once $\theta \neq 0$, there is mixing between the $F^{nA}$ and $\t F^{nA}$ multiplets, and the
boundary term \eqref{bry_var} implies the boundary conditions
\be\label{bry_eq}
F^{nA} = \tan(\alpha) \t F^{nA} - gJ^A \ , 
\qquad
\lambda_- = - \tan(\alpha) \lambda_+ + i g \left( \psi \phi^* - \psi^c \phi \right) \ , 
\ee
so as to have a well-defined variational problem. 
(One is free to impose the equation of motion $D=0$ to the auxiliary field above.)  

From a quantum or path integral point of view, the perspective on the boundary conditions shifts from varying the fields to summing over them.  The questions are whether and how to sum over boundary values of the fields.  Dirichlet boundary conditions are equivalent to treating the boundary values of the fields as fixed external sources.  Instead, we would like to treat the boundary values as dynamical and sum over them, allowing the bulk and boundary fields to interact with each other. 
Up to some numerical factors, integrating the bulk action by parts leads to the expression (\ref{bry_var}) but with $\delta A_A$ and $\delta \t \lambda_+$ replaced with $A_A$ and $\t \lambda_+$.  Morally, integrating over the boundary values of $A_A$ and $\t \lambda_+$ in the path integral then leads to Dirac delta like conditions enforcing the boundary conditions (\ref{bry_eq}).

As we work in perturbation theory, we should emphasize that our starting point is to find the free propagators in the bulk and on the boundary. We thus start with the boundary conditions
\be \label{N1_BC}
F^{nA} 
= \tan(\alpha) \t F^{nA}\ , 
\qquad 
\lambda_- = - \tan(\alpha) \lambda_+  \ ,
\ee
and interpret \eqref{bry_eq} as perturbative corrections to the free theory.

For the fermion we can put the boundary conditions in a more convenient form by using the definition $\lambda_- = \Pi_+ \gamma^5 \lambda$. The boundary condition is then equivalent to $\Pi_- e^{-\alpha \gamma^5} \lambda=0$.
We can define a new projection $\Pi^\theta_- = e^{\alpha\gamma^5} \Pi_- e^{-\alpha\gamma^5}$ such that the boundary condition is simply $\Pi_-^\theta \lambda=0$. The conjugation of $\Pi_-$ by $e^{\alpha\gamma^5}$ is clearly analogous to the action of the R-symmetry and has the effect of shifting $\eta \to \eta - 2 \alpha$ in $\Pi_-$. This shift is reminiscent of the relation between the $\theta$-term and an anomalous R-symmetry. To avoid confusion, let us emphasize that $\Pi_-^\theta$ is only relevant for the boundary conditions on $\lambda$. The preserved supercharge is still determined by $\Pi_+$ and so are the definitions of $\lambda_\pm$ which are the components of the multiplets of the preserved supersymmetry.

\subsection{${\mathcal N}=2$ Super Graphene}
\label{sec:N2}

In the case of $\cN=2$ there are two copies of the minimal supercharge $Q_i$ with $i=1,2$. The supercharges form a doublet of the $SU(2)_{\text{R}}$ symmetry. Since the R-symmetry is incompatible with the Majorana condition, the supercharges must instead admit a symplectic Majorana condition (see Appendix \ref{app:spinor})
\be
\bar Q_i = \varepsilon_{ij} Q^{jT} C_+  \ ,
\ee
where $C_+ = i \gamma^5 C$ is a new charge conjugation matrix obeying $C_+ \gamma^\mu C_+^{-1} = \gamma^{\mu \,T}$. The $\cN=2$ algebra is given by $\{ Q^i , \bar Q_j \} = 2i \delta^i_j \slashed{\partial}$. Repeating the analysis of the previous section we define projection operators 
\be
\label{PipmdefNtwo}
\Pi_\pm = \frac{1}{2} \left(1\pm \vec v \cdot \vec \tau \, \beta  \right) \ ,
\ee
 where $\vec v$ is a unit vector, $\vec\tau^{\,i}{}_j$ are the generators of the $SU(2)_{\rm R}$ 
 algebra normalized so that $(\vec v \cdot \vec\tau)^2=1$,
 and $\beta = i \gamma^n \gamma^5 e^{\eta\gamma^5}$ as before. 
 We choose the supercharges defined by the positive projector, namely $Q_+ = \Pi_+ Q$ and its complex conjugate. They generate a 3d $\cN=2$ subalgebra given by $\{ Q_+ , \t Q_+ \} = 2i \Gamma^A \del_A$. Previous definitions of the 3d bar $\t Q$, the gamma matrices $\t \gamma^A$ and so forth apply here without change. 

We are here treating $Q_+$ as a 3d Dirac spinor. Lest this cause any confusion, let us briefly comment on the apparent vanishing of the $SU(2)$ indices in the definition $\Pi_+ Q$. The matrix $\vec v \cdot \vec \tau$ has eigenvalues $\pm 1$. In the subspace of the $+1$ eigenvalue, the projector becomes $\frac{1}{2}(1+\beta)$, and acting on a 4-component spinor it gives the 3d Dirac spinor $Q_+$. In the subspace of $-1$ we get, up to multiplication by $\t C_+$ and transposition, the 3d Dirac spinor $\t Q_+$. We shall review the relation between 3d and 4d spinors in more detail later in this section. 

Let us consider the global symmetries more closely. In addition to the $SU(2)_{\rm{R}}$ mentioned before, the 4d $\cN=2$ algebra also has a $U(1)_{\rm R}$ symmetry acting on $Q^i$ by $e^{\eta \gamma^5} Q^i$. This symmetry is necessarily broken in the presence of a boundary, just as in the $\cN=1$ case, since $\gamma^5$ doesn't commute with the projection operator. In contrast, the $U(1)$ subgroup of $SU(2)_{\rm R}$ generated by $\vec v \cdot \vec \tau$ clearly commutes with the projector and corresponds to the R-symmetry of 3d $\cN=2$ acting on $Q_+$ by $e^{i\varphi} Q_+$ and in the $-1$ subspace by $e^{-i\varphi} \t Q_+$. Labelling the two other generators of $SU(2)_{\rm R}$ by $(\tau_\pm)^i{}_j$ with the anti-commutative property $\{\tau_\pm, \vec v \cdot \vec \tau \}=0$, we get the useful identity $\Pi_+ \tau_\pm = \tau_\pm \Pi_- $. As long as the R-symmetries are unbroken in the bulk we can use them to set $\eta=0$ and $\vec v$ to a convenient value, although we shall not bother to do that. 

We again write the total action as 
\be
\label{n2totaction}
S_{\rm tot} = S_{\rm bulk} + S_{\rm bry} \ .
\ee
The bulk now contains a photon $A_\mu$, 2 photinos $\lambda^i$, and 2 real scalars $S$ and $P$.  The photinos correspond to a pair of symplectic Majorana fermions.   
The bulk action reads
\be
S_{\rm bulk} = \int_{\cal M} \d^4 x \, \left( - \frac{1}{4} F_{\mu\nu} F^{\mu\nu} + \frac{i}{2} \bar \lambda_i \slashed{\partial} \lambda^i - \frac{1}{2} (\partial_\mu S)^2 -\frac{1}{2} (\partial_\mu P)^2  + \frac{1}{2} \vec D^2 \right) \ .
\ee
The vector $\vec D$ denotes 3 auxiliary fields.
We will introduce a $\theta$-term separately below to avoid clutter.

The boundary fields are the same as they were before -- a complex scalar $\phi$ and 3d Dirac spinor $\psi$ along with a complex auxiliary field $F$.  However, there are several additional Yukawa couplings involving the bulk fields: 
\be
\label{n2action}
S_{\rm bry} 
&=&  \int_{\partial {\cal M}} \d^3 x \, \Biggl(- \frac{1}{4} \bar \lambda_i \, \vec v \cdot \vec \tau^{\,i}{}_{j} \gamma^5 e^{\eta \gamma^5} \lambda^j
- X (\vec v \cdot \vec D + \partial_n X) 
 \nonumber \\
&& ~~~~~~~~~~~~~ + 
i\t \psi \slashed{D} \psi- |D_A \phi|^2   + |F|^2 + \sqrt{2} ig \big(\phi^* \, \t\lambda_+ \psi   -\phi \, \t \psi \lambda_+ \big) 
 \nonumber \\
&&
~~~~~~~~~~~~~ +g \t \psi \, Y \psi 
-g^2 |\phi|^2 Y^2 
-  g (\vec v \cdot \vec D + \partial_n X)|\phi|^2  \Biggr) \ ,
\ee 
where $ \lambda_+  =  \Pi_+ \lambda^1$ is taken to be a 3d Dirac spinor (more on that below), and the real 
scalars $X$ and $Y$ are defined by 
\be \label{PS_to_XY}
\begin{pmatrix}
  X \\
  Y \\
\end{pmatrix}
=
\begin{pmatrix} 
-\sin \eta & \cos \eta \\ -\cos \eta  & -\sin \eta
\end{pmatrix}
\begin{pmatrix}
  S \\
  P \\ 
\end{pmatrix}
\ . 
\ee  

The action \eqref{n2totaction} is invariant off-shell 
under the following SUSY transformations with the variation parameter satisfying $\Pi_+ \epsilon_i = \epsilon_i$\ :
\be
\delta A_\mu &=& -i\bar \epsilon_i \gamma_\mu \lambda^i \ , \\ 
\delta \lambda^i &=& \left(\frac{1}{2} \gamma^{\mu\nu} F_{\mu\nu} \delta^i_j  
- i\vec D \cdot  \vec \tau^{\,i}{}_{j} \right)\epsilon^j + \slashed{\partial} (S + \gamma^5 P) \epsilon^i 
\ , \\
\delta S &=& -i \bar \epsilon_i \lambda^i \ , \\ 
\delta P &=&- i \bar \epsilon_i \gamma^5 \lambda^i  \ , \\
\delta \vec D &=&   \vec \tau^{\,i}{}_j \bar \epsilon_i \slashed{\partial} \lambda^{j}  \ , 
\ee 
and
\be
\frac{1}{\sqrt{2}}\delta \phi &=& - \t \epsilon \, \psi \ , \\
\frac{1}{\sqrt{2}}\delta \psi &=& \Big( i  \Gamma^A \epsilon D_A \phi -  F \epsilon^c \Big)- g Y \epsilon \phi\ , \\
\frac{1}{\sqrt{2}}\delta F &=& \t \epsilon^{\, c} \,  \frac{\partial {\mathcal L}}{\partial \t \psi}
=\Big( i \, \t \epsilon^{\, c} \slashed{D} \psi -ig  \phi \t \epsilon^{\, c} \, \lambda_+ \Big)+g Y \t \epsilon^{\, c} \psi  \ ,
\ee  
where $ \epsilon = \epsilon^1$ and $\epsilon^c = - \t C_+^{-1} \t \epsilon^{\, \, T}$ (where $\t C_+ \equiv e^{\eta \gamma_5} C_+$ like in the ${\mathcal N}=1$ case).

To verify supersymmetry, 
it is instructive to derive the 3d multiplet of $A_A$ which participates in the boundary interactions. We find
\be \label{n2_3d_vec1}
\delta A_A &=& -i \, \t \epsilon \, \Gamma_A \lambda_+  + i \t \lambda_+ \Gamma_A \epsilon \ , 
\\
\delta \lambda_+ &=& \frac{1}{2} \Gamma^{AB} \epsilon F_{AB} -i \left(\vec v \cdot \vec D + \del_n X \right) \epsilon - \Gamma^A \epsilon \,\del_A Y \ ,
\\
\delta Y &=&   i \, \t \epsilon \, \lambda_+ - i \t \lambda_+ \epsilon  \ , 
\\ \label{n2_3d_vec4}
\delta \left( \vec v \cdot \vec D + \del_n X \right) &=&  \t \epsilon \, \Gamma^A \del_A \lambda_+ +  \del_A \t \lambda_+ \Gamma^A \epsilon \ .
\ee
The multiplet we have obtained is the 3d $\cN=2$ vector multiplet in Wess-Zumino gauge. In addition to the effective gauge field on the boundary $A_A$, this multiplet comprises a Dirac spinor $\lambda_+$, a real scalar $Y$ and a real auxiliary field $\vec v \cdot \vec D + \del_n X$. The appearance of this particular combination of the 4d fields as effective components of the boundary vector multiplet explains the form of the interactions in \eqref{n2action}.

Let us now consider adding a $\theta$-term. As in the $\cN=1$ preserving case, supersymmetry requires a compensating boundary action, which can easily be derived using the variation \eqref{n2_3d_vec1}. We find
\begin{align} \label{}
S_{\theta} = \int_{\mathcal M} \d^4 x \frac{g^2 \theta}{16\pi^2} F^{\mu\nu} \t F_{\mu\nu} - \int_{\del \mathcal M} \d^3 x \frac{g^2 \theta}{4\pi^2}\left(\t\lambda_+\lambda_+ - Y(\vec v \cdot \vec D + \del_n X ) \right) \ ,
\end{align}
As before we can absorb any Chern-Simons term on the boundary in this action. This way, the Chern-Simons level is the integer part of $\theta/2\pi$ which no longer has the symmetry $\theta \to \theta+2\pi$.

Let us now go on to discuss the details of the relation between the 4d Majorana spinors and the projected 3d Dirac spinors, and in doing so explain how the variations in \eqref{n2_3d_vec1}-\eqref{n2_3d_vec4} are derived. The symplectic Majorana condition can be written in a form adapted to the 3d subspace as
\be
\t \lambda_i = \varepsilon_{ij}  \lambda^{j\, T} \t C_+ \ , 
\ee
which gives us the relations  
\be
\lambda^2 = \t C_+^{-1} \t \lambda^T_1 \ , 
\qquad
\t \lambda_2 =  \lambda^{1\, T} \t C_+ \ . 
\ee
Without loss of generality we can associate $i=1$ with the $+1$ eigenvalue of $\vec v \cdot \vec \tau$ and $i=2$ with $-1$. We then have 
\be
(\Pi_+)^1{}_i \lambda^i = \frac{1}{2}(1+\beta) \lambda^1 \equiv \lambda_+ \ ,
\qquad
(\Pi_+)^2{}_i \lambda^i = \frac{1}{2}(1-\beta) \lambda^2 =   \t C_+^{-1} \t \lambda^T_+ \ .
\ee
Indeed, this is consistent since $\beta \t C_+^{-1} = - \t C^{-1}_+ \beta^T$. The required extra minus sign, relative to the analogous $\cN=1$ relation, which converts $(1-\beta)$ to $(1+\beta)^T$ is a result of using $C_+ = i \gamma^5 C$.  As an example consider 
\be
\delta A_A = -i \bar \epsilon_i \gamma_A \lambda^i = -i \, \t\epsilon_i \t \gamma_A \lambda^i \ .
\ee
Since $(\Pi_+)^i{}_j \epsilon^j = \epsilon^i$ we define $\epsilon^1 = \epsilon$ and the $i=1$ term immediately gives $-i \,\t\epsilon \,\Gamma_A \lambda_+$. For the $i=2$ term we have
\begin{align} \label{}
-i \, \t \epsilon_2 \t\gamma_A \lambda^2 
= 
-i \epsilon^{T} \t C_+ \t\gamma_A  \t C_+^{-1} \t \lambda_+^T
=
i\t\lambda_+ \Gamma_A \epsilon \ .
\end{align}

\subsubsection*{Boundary Conditions}

The arguments concerning boundary conditions are analogous to the ${\mathcal N}=1$ case.  The variational principle 
yields the same boundary condition for the gauge field as before, $F^{nA} = \frac{g^2 \theta}{4\pi^2} \t F^{nA} - g J^A$. 
Applying the equation of motion $\vec D = 0$ to the auxiliary fields, the boundary conditions for the scalars are  
\be
X - \tan \alpha Y &=&  g |\phi|^2 \ , \\
\partial_n (Y + \tan \alpha X) &=& -g \bar \psi \psi + g^2 |\phi|^2 Y \ .
\ee
This leads us to define new fields $X^\theta$ and $Y^\theta$ with
\be
\label{xthetaytheta}
\begin{pmatrix} X^\theta \\ Y^\theta \end{pmatrix}
=
\begin{pmatrix} \cos \alpha & - \sin \alpha \\ \sin \alpha & \cos\alpha \end{pmatrix}
\begin{pmatrix} X \\ Y \end{pmatrix} \ ,
\ee
such that the free-field boundary conditions are $X^\theta = 0$ and $\del_n Y^\theta = 0$. This new basis is analogous to $A^\theta_\mu$ introduced in the non-supersymmetric case \eqref{nonsusybrycond}, for which the corresponding boundary conditions are $F^\theta_{nA}=0$ for any $\theta$.
Because of its complexified nature in the ${\mathcal N}=2$ case, the photino boundary condition can be written in a
simpler form than before,
\be
\lambda_- + \frac{g^2 \theta}{4\pi^2} \lambda_+ = \sqrt{2} i g \psi \phi^* \ .
\ee
As in the $\cN=1$ case we can define new projections $\Pi_{\pm}^\theta = e^{\alpha\gamma^5} \Pi_{\pm} e^{-\alpha\gamma^5}$ such that the boundary conditions take the form $\Pi_-^\theta \lambda = \sqrt 2i g \cos \alpha \, \psi \phi^*$.

Morally, the ${\mathcal N}=4$ case corresponds to adding a couple more $X$, $Y$, and $\lambda$ fields to the ${\mathcal N}=2$ action.  To see a full $SU(4)$ R-symmetry along with its breaking pattern in the presence of a boundary is however more intricate.
We will omit a similar presentation of the ${\mathcal N}=4$ action and move on. As the general structure of the Yukawa interactions in the ${\mathcal N}=4$ case is already apparent from the ${\mathcal N}=2$ case we presented above, we do not need the details in the perturbative calculations to follow.  (See 
refs.\ \cite{Gaiotto:2008ak,Gaiotto:2008sd,Gaiotto:2008sa} for a discussion of boundary conditions for ${\mathcal N}=4$ Yang-Mills theory in 4d.)

%%%%%%%%%%%%%%%%%%%%%%%%%%
%%%%%%%%%%%%%%%%%%%%%%%%%%

\section{Propagators and 
Displacement Operator Correlators 
}
\label{sec:props}

In this section we will compute the leading order corrections to the anomaly coefficients $b_1$ and $b_2$ by slightly generalizing a free-field computation of these same coefficients.  
In general, the coefficients are related to two- and three-point functions of the displacement operator, which in turn is related to the boundary limit of the 
normal-normal component of the stress-tensor, $T_{nn}$. 
In the free-field limit, these correlation functions are straightforward to compute using Wick's theorem. 

As discussed in \cite{Herzog:2017xha}, the leading correction is completely captured by the one-loop 
self-energies of the bulk fields, where the self-energies come from interactions between the boundary limit of these bulk fields and the boundary degrees of freedom. 
Thus, we can obtain the corrected $b_1$ and $b_2$ by simply redoing the free-field computations but with resummed propagators that incorporate the one-loop self-energies.

We begin this section by discussing the propagators of the various bulk fields. We then use these propagators, along with self-energies to be obtained in section \ref{sec:oneloop}, to compute the corrected values of $b_1$ and $b_2$. 
The propagators are also important for performing the one loop computations in section \ref{sec:oneloop}.   

\subsection{Scalar}

In general the two-point function of a scalar field in the presence of a boundary is fixed by conformal symmetry up to a single function.\footnote{%
Without a boundary, the functional form of the two-point function is determined by the scaling dimension of the operators.  
In the presence of a boundary there is a non-trivial cross ratio \cite{JC, McAvity:1995zd,
McAvity:1993ue}. 
} 
In the present case, since the fields are free in the bulk, the two-point function is determined up to a choice of normalization and a reflection coefficient:
\be
\label{scalarG}
G_\Phi(x; x') &=&  \kappa_s \left(\frac{1}{|x-x'|^{d-2}}  + \Omega_{\rm{s}} \frac{1}{|\bar x - x'|^{d-2}} \right) \ , 
\label{scalarprop}
\ee 
where $x = (y, {\bf x})$ and $\bar x = (-y, \bf x)$ and $y=x^n$ 
is the normal direction.  
The conventional normalization is 
\be
\kappa_s = {1\over (d-2) \Vol(S^{d-1})} 
\ee 
with $\Vol(S^{d-1}) = 2 \pi^{d/2} / \Gamma(d/2)$.  
This form of \eqref{scalarprop} can be understood according to the method of images. The first term corresponds to the usual Green's function of the free bulk field equation, while the second is a homogeneous solution with an undetermined coefficient.
The variable $\Omega_{\rm{s}}$ determines the boundary conditions for bulk fields. 
Reflection positivity of $\langle \Phi(x) \Phi(x') \rangle$ along with $\langle \partial_n \Phi(x) \partial_n \Phi(x') \rangle$ further restrict $-1 \leq \Omega_{\rm{s}} \leq 1$.\footnote{%
 Ref.\ \cite{Herzog:2017xha}  noted these bounds in the context of the boundary conformal bootstrap program \cite{Liendo:2012hy}, 
for a particularly simple class of crossing equations associated with generalized free fields.
} 
Of all possible values of $\Omega_{\rm s}$ only those saturating the bound correspond to local boundary conditions on the fields, namely
\be \label{D_N_BC}
\Omega_{\rm s} 
= 
\begin{cases}
-1,  &\text{Dirichlet}, \\
1,  &\text{Neumann}.
\end{cases}
\ee
Other values of $\Omega_{\rm s}$ correspond to non-trivial boundary interactions. 
As explained in the previous section, in perturbation theory we use the propagator in the limit where the boundary interactions vanish. In particular, for the $\cN=2$ scalar fields $X$ and $Y$ we have the value $\Omega_{\rm s}=-1$ and $\Omega_{\rm s}=1$ respectively. We will see below that these initial values of $\Omega_{\rm s}$ get perturbative corrections.

A nonzero $\theta$ parameter alters the story somewhat by introducing mixing between the $X$ and $Y$ scalar fields.  However, we can always rotate to a frame (\ref{xthetaytheta}) where $X^\theta$ and $Y^\theta$ decouple and have the standard Dirichlet and Neumann boundary conditions in the $g \to 0$ limit.\footnote{Here we shall not distinguish between the conformal Robin boundary condition and the Neumann boundary condition as we focus on the flat limit.}

Since the interactions we consider are confined to the boundary, the internal propagators in a Feynman diagram are always boundary to boundary. It is thus useful to
work in a hybrid formalism where we replace the tangential coordinates ${\bf x}$ with momenta but leave the normal direction $y$ untouched.  Our convention for the Fourier transform is
\be
\label{scalarpropFT}
\tilde G_\Phi(y,y'; p) &=& \int \d^{d-1} {\bf x} \, e^{-i p \cdot {\bf x}} G_\Phi(y,{\bf x}; y', 0) = \frac{1}{2p} (e^{-p |y-y'|} +\Omega_{\rm{s}} e^{-p|y+y'|}) \ .
\ee
In this form it is easy to verify \eqref{D_N_BC}, e.g. the propagator $\tilde G_X$ restricted to the boundary $y=0$ vanishes, as it should given the Dirichlet boundary conditions.  

The propagator for $\tilde G_\Phi$ has one fewer power of $p$ in the denominator than the typical fully Fourier transformed propagator used in the absence of a boundary.  As the Feynman rules we use are the usual ones but with this alternate propagator, the changes in the physics we find can often be traced back to this change in the power of $p$ in the denominator of the propagator. 

Note that in Wick rotating the Fourier transforms to real time and converting them to conventional Feynman propagators, a factor of $1/i$ appears.

\subsection{Photon}

The photon in the free limit, for $\theta = 0$, satisfies either absolute or relative boundary conditions.  By the method of images, in the Feynman gauge, the correlation functions are given by 
\be
\label{photonpropone}
G_{nn}(x; x') &=& \kappa_s \left(\frac{1}{|x-x'|^{d-2}}  - \Omega_{\rm v} \frac{1}{|\bar x - x'|^{d-2}} \right)\ , \\
G_{AB}(x; x') &=&  \kappa_s \left(\frac{1}{|x-x'|^{d-2}}  + \Omega_{\rm v} \frac{1}{|\bar x - x'|^{d-2}} \right) \eta_{AB} \ ,
\label{photonproptwo}
\ee
and their Fourier transforms can be read off from the result for the scalar fields, 
\be
\label{photonpropFTone}
\tilde G_{nn}(y,y'; p) &=& \frac{1}{2p} \big( e^{-p |y-y'|}  - \Omega_{\rm v} e^{-p|y+y'|}\big) \ , \\
\tilde G_{AB}(y,y'; p) &=&  \frac{1}{2p} \big( e^{-p |y-y'|} + \Omega_{\rm v} e^{-p|y+y'|}\big) \eta_{AB} \ .
\label{photonpropFTtwo}
\ee
The absolute boundary conditions of interest in this paper, which preserve a nonzero boundary value of $F_{AB}$, 
correspond to $\Omega_{\rm v} = 1$.  The relative choice, more familiar from electrostatics problems where the boundary is an equipotential surface, is $\Omega_{\rm v} = -1$.

To incorporate a $\theta \neq 0$, one can interpret the propagators (\ref{photonpropone}) and (\ref{photonproptwo}) as those of the ``rotated'' photon $A^\theta_\mu$.  The correlation functions for the original field strengths can then be extracted from the boundary condition relation $F^\theta_{\mu\nu} = \cos(\alpha) F_{\mu\nu} - \sin(\alpha) \t F_{\mu\nu}$ and a corresponding equality for $\t F^\theta_{\mu\nu}$.  
Wick rotating and converting to Feynman propagators, we need again a factor of $1/i$.

\subsection{Photino}

We can use arguments similar to those made for the scalar to constrain the bulk fermion two-point function. This leads to% 
\footnote{%
We use this opportunity to correct the overall sign typo in the fermion propagator in \cite{ Herzog:2017kkj,Herzog:2017xha}. The final results in these papers are not changed.
}
\be
\label{Gphoti}
G_{\lambda}(x; x') &=& -\kappa_f \left( \frac{i \gamma \cdot (x - x')}{|x - x'|^d} + \Omega_{\rm f} \frac{i \gamma \cdot (\bar x - x')}{|\bar x - x'|^d }\right)  ,
\label{lambdaprop}
\ee
where $\kappa_f = 1 / \Vol(S^{d-1})$. It is straightforward to check that the image term satisfies the Dirac equation acting from the right on $x'$. Since the same must be true for the Dirac operator acting on $x$ from the left we must have that 
$\Omega_{\rm f} \, \gamma \cdot \bar x = \gamma \cdot x \, \Omega'_{\rm f}$ 
for some $\Omega'_{\rm f}$. 
In fact we can show that $\Omega'_{\rm f} = \bar \Omega_{\rm f}$.  
The  two-point function $\langle \lambda(x) \bar \lambda (x') \rangle = \overline{\langle \lambda(x') \bar \lambda (x) \rangle}$ must be self-conjugate,  which in turn implies that $\Omega_{\rm f} \, \gamma \cdot \bar x = \gamma \cdot x \, \bar\Omega_{\rm f}$.  In components, this relation is the already familiar
\be \label{Omega_cond}
\Omega_{\rm f} \gamma^n + \gamma^n \bar\Omega_{\rm f} = 0\ ,\qquad
\Omega_{\rm f} \gamma^A - \gamma^A \bar\Omega_{\rm f} = 0\ .
\ee 
These equations are precisely the same we found for $\beta=i\gamma^n\gamma^5 e^{\eta\gamma^5}$  except that $\Omega_{\rm f}$ does not necessarily square to one in 
interacting theories. Moreover, the phase of $\Omega_{\rm f}$ does not have to be correlated with $\eta$, the phase of $\beta$ which is determined by the preserved subalgebra, i.e.\ the phase of $\Pi_+$. In fact, as explained below \eqref{N1_BC}, such a relative phase is a consequence of a $\theta$-term. The change in the fermion boundary condition coming from $\theta$ leads to $\beta^\theta = i \gamma^n \gamma^5 e^{(\eta - 2\alpha)\gamma^5}$ with $\tan \alpha =  \frac{g^2 \theta}{4\pi^2}$. Putting everything together, we will see below that supersymmetry fixes the form to  $\Omega_{\rm f} = \Omega_{\rm v} \beta^\theta $ for $\cN=1$ and $\Omega_{\rm f} = \Omega_{\rm v}(\vec v \cdot \vec \tau) \beta^\theta $ for $\cN=2$.

As in the case of the scalars, in the context of perturbation theory we consider the free propagator with $\Omega_{\rm f} = \beta$ for $\cN=1$ or $\Omega_{\rm f} = (\vec v \cdot \vec \tau)\beta$ for $\cN=2$, setting $\theta=0$ for simplicity.%
\footnote{Alternatively, the results for the propagator apply without change provided we make the substitutions $\t \gamma^A \to e^{2\alpha\gamma^5} \t\gamma^A$, $\t\lambda \to \t\lambda e^{-2\alpha\gamma^5}$ etc. These substitutions are of course just a shift $\eta \to \eta-2\alpha$ in the definitions $\t\gamma^A$ and $\t \lambda$.} Focusing for simplicity on the $\cN=1$ case,
the Fourier transform of \eqref{Gphoti} is then
\be
\label{lambdapropFT}
&&\tilde G_\lambda(y,y'; p) = \int \d^{d-1} {\bf x} \, e^{-i p \cdot {\bf x}} G_\lambda(y,{\bf x} ;y', 0) \nonumber \\
&=& -\frac{1}{2} \left(  \frac{ \gamma^A p_A}{p}+ i \sgn(y-y') \gamma^n \right) e^{-p |y-y'|} %+
- \frac{\beta}{2} \left(  \frac{ \gamma^A p_A}{p}- i \gamma^n \right) e^{-p (y+y')} \ .
\ee
To get more insight into the propagator it is convenient to use the language adapted to 3d. To this end we rewrite the two-point function as $\langle \lambda(x) \t \lambda(x') \rangle$ which has the effect of substituting $\gamma^\mu \to \t\gamma^\mu$ in \eqref{lambdapropFT}. Consider taking one of the insertion points $x$ to the boundary, i.e. $y=0$. We find
\be
\label{lambda_two_point}
\langle \lambda(0; p) \t\lambda(y' ; -p) \rangle = - \Pi_+ \left( \frac{\t\gamma^A p_A}{p} - i \t \gamma^n \right) e^{-y' p} \ .
\ee
Acting with $\Pi_-$ from the left clearly annihilates this expression which is a reflection of the boundary condition $\lambda_-=0$. A similar expression is found for $y'=0$ with $\Pi_+$ appearing to the right of the brackets in \eqref{lambda_two_point} thus reflecting the boundary condition of $\t\lambda$. As before, we only encounter boundary to boundary propagators with $y=y'=0$ in which it is clear from \eqref{Gphoti} that there is no $\gamma^n$ term. (The result in \eqref{lambdapropFT} is obtained based on the assumption that either $y$ or $y'$ are non-vanishing.) The propagator then becomes 3d 
\be
\langle \lambda(0; p) \t\lambda(0 ; -p) \rangle = - \Pi_+ \frac{\t\gamma^A p_A}{p}  = - \frac{\Gamma^A p_A}{p} \ .
\ee

In Wick rotating to real time and converting to Feynman propagators, again a factor of $1/i$ appears.

%%%%%%%%%%%%%%
%%%%%%%%%%%%%%
\subsection{Relations between $\Omega_{\rm s}$, $\Omega_{\rm v}$ and $\Omega_{\rm f}$}

Let us now show how supersymmetry relates $\Omega_{\rm s}$, $\Omega_ {\rm v}$ and $\Omega_{\rm f}$. 
To facilitate the comparison let us denote $\Delta(x-x') = \kappa_s / | x- x' |^{d-2}$, such that
\be
G_\Phi = \Delta(x-x') + \Delta(x - \bar x') \Omega_{\rm s} \ , 
\qquad
G_\lambda = i \slashed{\del} \Big( \Delta(x-x') + \Delta(x - \bar x') \Omega_{\rm f}\Big) \ ,
\ee
and likewise for the gauge field.  
We have written the fermion propagator in the tilde frame, so by $\slashed{\del}$ we here mean $\t \gamma^\mu \del_\mu$. Note in addition that $\Omega_{\rm f}$ is self-conjugate in this frame, which means $\Omega_{\rm f} \,\t\gamma \cdot \bar x = \t\gamma \cdot x \, \Omega_{\rm f}$.

To relate the propagators, we use the fact that supersymmetry transformations associated with the preserved subalgebra annihilate the vacuum, and therefore correlation functions of expressions which are exact supersymmetry variations vanish.  Consider first, in the $\cN=1$ case, the multiplet $(A_A, \lambda_+)$ whose variations are found in \eqref{lambda_A_mult}. This leads to
\be
0 
&=& 
\langle \delta \left( A_A(x) \t\lambda_+(x') \right) \rangle\nn\\
&=&
-i \t\epsilon\, \t\gamma_A \langle \lambda_+(x) \t\lambda_+ (x') \rangle
- \frac{1}{2} \t\epsilon \,\t\gamma^{BC} \langle A_A(x) F_{BC}(x') \rangle \ ,
\ee
which, modulo a gauge transformation, gives the relation $\Pi_+ \Omega_{\rm f} = \Pi_+ \Omega_{\rm v}$ and implies%\footnote{The projector $\Pi_+$ basically neutralizes all of the matrices in $\Omega_f$ and picks out the degrees of freedom which have eigenvalue one.} 
\be
\label{Rn1}
\cN=1:~~~
\Omega_{\rm f} = \Omega_{\rm v} \beta \ . 
\ee
An almost identical derivation gives the same relation for $\cN=2$. To find the relation between the $\cN=2$ scalars and fermions we look at the correlation function
\be
0 
&=& 
\langle \delta \left( Y(x) \t\lambda_+(x') \right) \rangle\nn\\
&=&
i \t \epsilon \, \langle \lambda_+(x) \t\lambda_+(x') \rangle - \t\epsilon \, \t\gamma^A \langle Y(x) \del_A Y(x') \rangle \ ,
\ee 
which gives the relation 
\be
\label{fY}
\cN=2: ~~~\Omega_{\rm f} = \Omega_{Y} (\vec v \cdot \vec \tau)\beta \ .
\ee 
Using the transformations 
\be
\delta \lambda_- = \t \gamma^A \epsilon \del_A X + \ldots \ ,
\qquad
\delta X = -i \,\t\epsilon \,\lambda_- + i \t\lambda_- \epsilon \ ,
\ee
where the ellipses correspond to terms with fields other than $X$, a similar correlation function with $X$ and $\t\lambda_-$ gives 
\be
\label{Rn2}
\cN=2: ~~~\Omega_{\rm f} = -\Omega_{X} (\vec v \cdot \vec \tau)\beta \ . 
\ee
To see where the extra sign comes from, recall the definition $\lambda_- = \gamma^5 \Pi_- \lambda = \Pi_+ \gamma^5 \lambda$. The quantity $\langle \lambda_-(x) \t \lambda_-(x') \rangle$ is thus proportional to 
\begin{align} \label{}
\gamma^5 G_\lambda \gamma^5 = i \slashed{\del} \Big( \Delta(x-x') - \Delta(x - \bar x') \Omega_{\rm f}\Big) \ .
\end{align}
Given these relations between the $\Omega_X$, $\Omega_Y$, $\Omega_{\rm f}$, and $\Omega_{\rm s}$, let us introduce the following universal scaling factor:
\be
\label{R3}
\Omega = \Omega_Y = - \Omega_X= \Omega_{\rm v}  \ .
\ee

\subsubsection*{Adding Interactions}

In the previous section, we analyzed the propagator for general values of $\Omega$ and explained that in the free limit $|\Omega | = 1$.  
Now let us consider how the self-energies lead to a modification of the boundary condition parameter $\Omega$.
In the next section, we will see that the one-loop self-energies of the fields take the following form:
\be
\tilde \Pi_{(X)}(p) &=& \sigma (g) p  \ , ~~ \tilde \Pi_{(Y)}(p) = - \sigma(g) p  \ , \\
\tilde \Pi_{(\lambda)}(p) &=& \sigma(g) \frac{\slashed{p}}{p} \ , ~~
\tilde \Pi^{AB}_{(A_\mu)}(p) = - \frac{\sigma(g)}{p} (p^2 \eta^{AB} - p^A p^B) \ .
\ee 
It turns out that supersymmetry guarantees the 
function $\sigma(g)$ showing up in each of these self-energies is the same:  
we find\footnote{Removing only fermions from our actions would lead to certain mixed dimensional scalar QED type theories 
which still have boundary interactions and, presumably, the corresponding $\sigma$ could depend on the coupling. 
In the supersymmetric cases we are interested in here, sending $N_f \to 0$ implies removing scalars as well and the theories become free.}
\be
\sigma = \frac{(g \cos \alpha )^2}{8} N_f \ ,
\ee
where $N_f$ counts the number of Dirac fermions propagating on the boundary.
(To trust perturbation theory, the quantity $(g \cos \alpha)^2 N_f$ should be kept small.)

Having the above result, we can determine how $\Omega$ depends on $\sigma(g)$. 
Let us take the $Y$ field as the simplest example.
Concatenating the self-energy with scalar propagators to the boundary 
and away from the boundary yields the shift in the two-point function:
\be
\delta \tilde G_Y(y,y'; p) &=& \tilde G_{Y} (y,0;p) \tilde \Pi_{(Y)}(p) \tilde G_Y (0,y'; p) \nn\\
&=& -\frac{\sigma_Y e^{-p(y+y')} }{p} \ .
\ee  
Comparing with the Fourier transformed propagator (\ref{scalarpropFT}), we conclude that 
there is a perturbative shift in the boundary condition: 
\be
\label{sy}
\Omega = \Omega_Y = 1 - 2\sigma_Y + {\cal O}(g^4) \ .
\ee  
Similar computations for the remaining three fields 
-- $X$, $\lambda$, and $A_\mu$ -- 
yield 
results that are consistent with \eqref{Rn1}, \eqref{fY}, \eqref{Rn2}, and \eqref{R3}.
We note that the sign of the correction to $\Omega$ 
is consistent with the reflection positivity bounds $-1 \leq \Omega \leq 1$.

\subsection{Displacement Operator Two- and Three-Point Functions}

We are interested in the displacement two- and three-point functions because of their relation to the boundary anomaly coefficients $b_1$ and $b_2$, established by two of us in refs.\ \cite{Herzog:2017xha,Herzog:2017kkj}.  
For a small but nonzero value of the interaction $g$, the leading ${\cal O}(g^2)$ correction to these correlation functions comes from a modification of the boundary condition parameter $\Omega$ in the propagators.  

We will first review results for the displacement correlation functions computed from the free propagators using Wick's Theorem. 
Given the form of the bulk propagators discussed above, we will then compute the leading ${\cal O}(g^2)$ correction to the displacement correlation functions by a slight generalization of the free-field computation.

In a boundary CFT, a central role is played by the displacement operator, which can be defined as a failure of the stress tensor conservation on the boundary: 
\be
\partial_\mu T^{\mu n}(x) = D^n({\bf x}) \delta(x^n) \ .
\ee
An integrated version of this definition relates the displacement operator to the boundary limit of the normal-normal component of the stress tensor:
\be
T^{nn}|_{\partial {\cal {M}}} = D^n({\bf x}) \ .
\ee
Conformal invariance on the boundary constrains the form of the two- and three-point correlation functions up to constants, which we call $c^{nn}$ and $c^{nnn}$:
\be
\langle D^n({\bf x}) D^n(0) \rangle = \frac{c^{nn}}{|{\bf x}|^8} \ , \; \; \;
\langle D^n({\bf x}) D^n({\bf x}') D^n(0) \rangle = \frac{c^{nnn}} {|{\bf x}|^4 |{\bf x}'|^4 |{\bf x} - {\bf x}'|^4} \ .
\ee
Refs.\ \cite{Herzog:2017xha,Herzog:2017kkj} identified that
\be
b_1 = \frac{2 \pi^6}{35} c^{nnn} \ , \; \; \; b_2 = \frac{2 \pi^4}{15} c^{nn} \ .
\ee

\subsubsection*{Free Theories}

For free theories, we can calculate $c^{nn}$ and $c^{nnn}$ using Wick's theorem.  
The normal-normal component of the free-field stress tensor is
\be
T^{\Phi}_{nn} &=& (\partial_n \Phi)^2 - \frac{1}{12} ( 2 \partial_n^2 + \Box) \Phi^2 \ , \\
T^{\lambda}_{nn} &=& \frac{i}{4} \left( (\partial_n \bar \lambda) \gamma_n \lambda - \bar \lambda \gamma_n \partial_n \lambda \right) \ , \\
T^{A_\mu}_{nn} &=& \frac{1}{2} F_{nA} {F_n}^A - \frac{1}{4} F_{AB} F^{AB} \ , 
\ee for a scalar, a Majorana fermion and a gauge field, respectively. 
The results are described in greater detail in refs. \cite{McAvity:1993ue,Herzog:2017xha,Herzog:2017kkj}.  
Below we quote the 4d results in the case of interest. 

For a single real scalar with propagator (\ref{scalarprop}), 
one finds
\be
c^{nn}_\Phi &=& \frac{1}{4 \pi^4} \left( 1 + \Omega^2  \right) \ , \\
 c^{nnn}_\Phi &=& \frac{1}{36 \pi^6} (8 - 3 \Omega + 24 \Omega^2 - \Omega^3) \ .
\ee
For a single Majorana fermion with propagator (\ref{lambdaprop}) 
and four-dimensional gamma matrices
one finds  
\be 
c^{nn}_\lambda &=& \frac{3}{4 \pi^4} \left( 1 + \Omega^2 \right) \ , \\
c^{nnn}_\lambda &=& \frac{5}{8 \pi^6} (1 + 3 \Omega^2)
\ . 
\ee
For a photon with propagators (\ref{photonpropone}) and (\ref{photonproptwo}), 
one obtains 
\be
c^{nn}_{A_\mu} &=& \frac{3}{\pi^4} \left( 1+ \Omega^2  \right) \ , \\
 c^{nnn}_{A_\mu} &=& \frac{2}{ \pi^6} (1 + 3 \Omega^2) \ .
\ee 
In free theories, $\Omega^2=1$ and only the central charge $b_1$ of a scalar depends on boundary conditions. 

\subsection*{Adding Interactions}

While we computed these coefficients using Wick's Theorem and assuming free field theory, in the interacting case we claim that the ${\cal O}(g^2)$ correction is captured correctly by making the substitution $\Omega = 1 - 2 \sigma$ in these formulae.  At order $g^2$, the only diagrams that contribute to the stress-tensor two-point function are the free field diagrams and the one loop self-energy corrections.  (Starting at ${\cal O}(g^4)$, there are diagrams that involve scattering of four bulk fields.)

Note that $c^{nn} \sim (1- 2 \sigma)$ while $c^{nnn} \sim (1 - 3 \sigma)$.  
Without doing any further calculations, we see immediately that $b_1$ and $b_2$ must depend on the gauge coupling, regardless of the amount of supersymmetry.  Every one of our bulk fields leads to a reduction in $b_1$ and $b_2$ by an amount proportional to $\sigma$, and there is no possibility of cancellation.
It is tempting to conjecture that boundary interactions generally will never increase the values of these boundary central changes in a boundary CFT.  It would be interesting to search for a general argument or find a counterexample.
 
Let us 
explicitly calculate the corrections in the various cases.  
In the ${\mathcal N}=1$ case, including contributions from the photon and the photino we obtain
\be
b_1^{({\cal N}=1)} 
&=& \frac{2 \pi^6}{35} (c^{nnn}_{A_\mu} + c^{nnn}_\lambda) \nn\\
&=& \frac{3}{5} - \frac{9 g^2 N_f}{40}
+ {\cal O}(g^4)  \ , \\
b_2^{({\cal N}=1)} &=& \frac{2 \pi^4}{15} (c^{nn}_{A_\mu} + c^{nn}_\lambda) \nn\\
&=& 1 - \frac{g^2 N_f}{4}
+ {\cal O}(g^4) \ .
\ee
Here and below, we give the results when $\theta = 0$.  A $ \theta \neq 0$ can be restored by simply making the replacement $g \to g \cos(\alpha)$. 
In the ${\mathcal N}=2$ case, including the scalars and a second photino we obtain
\be
b_1^{({\cal N}=2)} &=& \frac{2 \pi^6}{35} (c^{nnn}_{A_\mu} + 2c^{nnn}_\lambda + c_X^{nnn} + c_Y^{nnn}) \nn\\
&=& \frac{38}{45}- \frac{19 g^2 N_f}{60}
+ {\cal O}(g^4)   \ , \\
b_2^{({\cal N}=2)} &=& \frac{2 \pi^4}{15} (c^{nn}_{A_\mu} + 2c^{nn}_\lambda + c_X^{nn} + c_Y^{nn})\nn\\
&=& 
\frac{4}{3}- \frac{g^2 N_f}{3}
+ {\cal O}(g^4) \ .
\ee
In the ${\mathcal N}=4$ case,  since the interactions between the additional bulk fields and the boundary matter should simply be duplicates of the interactions we have already studied, the result can be deduced by including an extra couple of photinos and bulk scalars.  We obtain
 \be
 b_1^{({\cal N}=4)} 
&=& \frac{2 \pi^6}{35} (c^{nnn}_{A_\mu} + 4c^{nnn}_\lambda + 3c_X^{nnn} + 3c_Y^{nnn}) \nn\\
&=& 
\frac{4}{3} - \frac{g^2 N_f}{2}
+ {\cal O}(g^4)   \ , \\
b_2^{({\cal N}=4)} &=& \frac{2 \pi^4}{15} (c^{nn}_{A_\mu} + 4c^{nn}_\lambda + 3c_X^{nn} + 3c_Y^{nn}) \nn\\
&=& 
2 - \frac{g^2 N_f}{2}
+ {\cal O}(g^4) \ .
\ee 
In all cases, we find that the first-order correction to these anomaly coefficients is nonzero.  In other words, these coefficients depend on the marginal coupling $g$.
(The zero-th order contributions of these results correspond to the boundary central charges in free theories.) 

Observe that the following quantity:
\be
\Delta b\equiv b_1 - b_2
\ee 
does not depend on $g$ in the ${\cal N}=4$ case (at leading order), suggesting the combination of curvature invariants $\tr \hat K^3 \pm h^{\mu\nu} \hat K^{\rho \sigma}W_{\mu\rho  \nu\sigma}$ may play a special role in these types of boundary conformal field theories with ${\mathcal N}=4$ supersymmetry in the bulk.\footnote{In this case, the bulk charges are the same: $a=c={1\over 4}$.}  
It will be interesting to see if this quantity remains $g$-independent at higher orders.

\section{Perturbation Theory for Super Graphene}
\label{sec:oneloop}

\subsection{Renormalization Group Analysis}

In this subsection, we will argue that our supersymmetric graphene theories are examples of boundary conformal field theory, with an exactly marginal coupling -- the gauge coupling -- to all orders in perturbation theory.  This discussion however neglects two issues that are worth further scrutiny but will not be discussed here.  
The first is the
possibility of non-perturbative contributions to the beta function.  
While instantons should be absent in the abelian gauge theory, it is not obvious how magnetic monopoles on the boundary might alter our theories.
The second is stability, for example spontaneous breaking of the $U(N_f)$ flavor symmetry.  We believe that for sufficiently small coupling, the theory should be stable  \cite{Gorbar:2001qt,Kaplan:2009kr,Kotikov:2016yrn}, but this issue and the other one deserve further consideration.

Let us begin with a discussion of the superficial degree of divergence of the different diagrams.  Topological constraints  along with momentum conservation imply that the superficial degree of divergence of an arbitrary diagram, regardless of loop level, is
\be
\label{sdd}
\frac{1}{2} \left( 6 - 2 n_A - 4 n_X - 2 n_Y - 3 n_\lambda - n_\phi - 2 n_\psi \right) \ .
\ee 
The quantity $n_\Phi$ is the number of external legs of the field $\Phi$.  
The reason that $n_X$ and $n_Y$ have different coefficients is that they have different boundary conditions.  An external $X$ leg can only couple to the diagram through a $|\phi|^2 \partial_n X$ vertex.  That restriction in turn means the $\partial_n$ must produce a power of an external momentum  that is not integrated over in the diagram and thus does not contribute to a short distance divergence.

Another useful quantity to consider is the power of the loop momenta, modulo two, in the numerator of an arbitrary diagram.  This power is
\be
\label{numpow}
n_A + n_Y + n_\phi + \frac{1}{2} (n_\lambda + n_\psi) \ .
\ee
If this power is odd, then by rotational invariance, the leading divergence of the diagram is reduced by one.  

The expressions (\ref{sdd}) and (\ref{numpow}) are useful for deducing a number features of the one loop calculations we will perform in the next subsection and also for extrapolating those results to arbitrary loop order.  Let us begin with the self-energy of the bulk fields.  These self-energies must all be finite, as one can see in a variety of ways.  The easiest is perhaps locality:  boundary interactions cannot renormalize the photon, photino, or $X$ and $Y$ scalar wave functions.  Indeed, we will see this finiteness explicitly at one loop in the next subsection.  But the naive power counting implicit in eqs.\ (\ref{sdd}) and (\ref{numpow}) largely bears out these observations as well.  

Consider first the self-energy of the photon, for which $n_A = 2$ with all the other $n_\Phi$ set to zero.  We see immediately that the diagrams should have superficial degree of divergence 1.  However, the photon self-energy must be accompanied by a gauge invariant prefactor $q^\mu q^\nu - g^{\mu\nu} q^2$ which immediately cuts down the degree of divergence by 2, rendering these diagrams finite.

Given the result for the photon, supersymmetry can be used to argue that the photino as well as the $X$ and $Y$ fields have no wave function renormalization.  Let us nevertheless repeat the naive power counting arguments.
For the photino self-energy, eq.\ (\ref{sdd}) suggests the diagrams are logarithmically divergent.  However, from eq.\ (\ref{numpow}), rotational invariance cuts down the degree of divergence by one.
 Based on eq.\ (\ref{sdd}), the self-energy of $X$ should be finite. 
 The $Y$ field at last provides an example where eqs.\ (\ref{sdd}) and (\ref{numpow}) are insufficient to give the right answer. 
 Naively, $Y$ should be linearly divergent.  However, what happens at least at one loop level is that two diagrams contribute, and their divergences cancel.

The self-energies of the boundary degrees of freedom and the boundary vertices are less well behaved.  
For the most part, they all have log divergences and corresponding wave-function renormalization.  
The eqs.\ (\ref{sdd}) and (\ref{numpow}) are sufficient to give the correct log divergence for the electron self-energy as well as the $\bar \psi \psi A_\mu$, $A_\mu |\phi|^2$, and $A_\mu^2 |\phi|^2$ vertices.  While eqs.\ (\ref{sdd}) and (\ref{numpow}) are insufficient to see it, through supersymmetry, the selectron self-energy must be log divergent as well.  While eqs.\ (\ref{sdd}) and (\ref{numpow}) predict a log divergence for the $\lambda \psi \phi$ Yukawa vertex, what happens at least at one loop is that two diagrams contribute and the divergence cancels.  It would be interesting to see whether the finiteness is accidental or comes from some symmetry and persists at higher loop level.

Let us summarize our one loop results for the wave-function renormalizations.
From the one loop self-energies and vertex functions we compute in the next subsection, we can read off the various wave-function renormalization $Z$-factors in the Lagrangian.\footnote{We follow the conventions of Srednicki's field theory text book \cite{Srednicki}.  As before, a nonzero $\theta$ can be incorporated by making the replacement $g \to g \cos(\alpha)$.}
We will divide the singular terms into two contributions. The first involves loops without a photino, and the second comes from loops with a photino. 
In the ${\mathcal N}=1$ case, the $Z$-factors are 
\be
Z_\psi &=&  1 + g^2 \left( - \frac{1}{6 \pi^2 \epsilon} - \frac{1}{3 \pi^2 \epsilon} + {\rm finite} \right)  = 1 + g^2 \left( - \frac{1}{2 \pi^2 \epsilon} + {\rm finite} \right) \ , \\
Z_\phi &=& 1 + g^2 \left( \frac{5}{6 \pi^2 \epsilon} - \frac{1}{3 \pi^2 \epsilon} + {\rm finite} \right) =  1 + g^2 \left( \frac{1}{2 \pi^2 \epsilon} + {\rm finite} \right) \ , \\
Z_{A_\mu \psi \psi} &=& 1 + g^2 \left( - \frac{1}{6 \pi^2 \epsilon} - \frac{1}{3\pi^2 \epsilon} + {\rm finite} \right)=Z_\psi \ , \\
Z_{A_\mu \phi \phi} &=& 1 + g^2 \left( \frac{5}{6\pi^2 \epsilon} - \frac{1}{3 \pi^2 \epsilon} + {\rm finite} \right)=Z_\phi \ ,
\ee
and
\be
Z_{\lambda \phi \psi} &=&1 +  g^2 \left(  \frac{1}{2\pi^2 \epsilon} - \frac{1}{2 \pi^2 \epsilon}+ {\rm finite} \right)= 1 + g^2 ({\rm finite}) \ , \\
Z_{A_\mu} &=& 1 + g^2 ({\rm finite}) \ , \\
Z_\lambda &=& 1 + g^2 ( {\rm finite}) \ ,
\ee where $\epsilon=4-d$ and the log divergences can be associated with the $1/\epsilon$ terms.
Note that the Ward identities which follow from gauge symmetry imply in a minimal subtraction scheme, to all loops, that $Z_\psi = Z_{A_\mu \psi \psi}$ and that $Z_\phi = Z_{A_\mu \phi \phi}$.  
Indeed, these Ward identities continue to be satisfied even if one removes the photino from the spectrum.  

The constraints from having a supersymmetric action mean that, to all loops, 
\be
\label{ZZ}
Z_\psi Z_\phi = Z_{\lambda \phi \psi}^2 \ . 
\ee  
While from the naive power counting discussed earlier we expect \eqref{ZZ} is log divergent order by order in perturbation theory, we have found something stronger at one loop, namely that $Z_{\lambda \phi \psi}$ and $Z_\psi Z_\phi$ are individually finite.  
We do not know if they remain individually finite at higher loop order.

The supersymmetric and gauge symmetry constraints on the $Z$-factors along with the finiteness of the $Z$-factors for the bulk fields $A_\mu$ and $\lambda$ are enough to guarantee that the gauge coupling is not renormalized at any loop order in perturbation theory for the ${\mathcal N}=1$ theory.  
For example, for the $A_\mu \psi \psi$ vertex, 
the relation between the bare coupling and physical coupling is given by $g_0 Z_{A_\mu}^{1/2} Z_\psi = g Z_{A_\mu \psi \psi}$, which then guarantees $g$ is independent of scale. 
Given the general arguments in this subsection, we did not actually need the detailed one loop results although they provide a useful check.

The same is true for the ${\mathcal N}=2$ theory.
In this case, we have $Z$-factors associated with the additional Yukawa interactions.  The structure of the Lagrangian and supersymmetry imply the following wave-function renormalization relations, to all loops,
\be
Z_{Y \psi^2} &=& Z_\psi \ , \\
Z_{X \phi^2}  &=& Z_{Y^2 \phi^2}  = Z_\phi  \ .
\ee
These relations, along with the finiteness of $Z_X$, $Z_Y$, $Z_\lambda$, and $Z_{A_\mu}$ are enough to guarantee that the beta functions for all of the Yukawa couplings vanish, without doing any one-loop calculations.  (The loop computations are still needed in order to determine the values of $b_1$ and $b_2$.)

The story for the ${\mathcal N}=4$ theory is very similar to what we just 
discussed above in the ${\mathcal N}=2$ case, the main difference being that we have more photinos and $X$ and $Y$ type fields at our disposal.  
The nature of the interaction vertices and propagators is the same, and the power counting arguments and Ward identities are analogous.  
Thus, we expect that the ${\mathcal N}=4$ theory also has a vanishing beta function for the gauge coupling at all orders in perturbation theory. 

Finally, let us discuss terms that do not appear in the Lagrangian, of the schematic form $XY$, $\psi^2 \phi^2$, and $\phi^6$, but which could in principle be generated at loop level.  (The vertex $\psi^2 \phi$ does not conserve charge and is related by supersymmetry to $\phi^4$.)

Rotational invariance (\ref{numpow}) means all of the $XY$ mixing diagrams are finite.  
The $\phi^2 \psi^2$ and $\phi^6$ couplings are the most interesting.  
They are classically marginal and related by supersymmetry via a $\Phi^4$ type superfield in the Lagrangian, were we to include it.\footnote{%
Note that we can arrange for charge conservation by having superfields $\Phi^+$ and $\Phi^-$ with opposite charges in a theory with $N_f>1$.
}

The constraints of eqs.\ (\ref{sdd}) and (\ref{numpow}) on $\psi^2 \phi^2$ are surprisingly stringent.  Naively, the diagram is log divergent, but rotational invariance makes it finite.  Thus, we expect the beta function for the $\psi^2 \phi^2$ to vanish at the point when the physical coupling itself vanishes.  
By supersymmetry, the story must be the same for the $\phi^6$ coupling although one cannot see it from eqs.\ (\ref{sdd}) and (\ref{numpow}).

The story changes somewhat if we include a bare $\psi^2 \phi^2$ coupling. The numerator of the loop integral will include an extra number of momenta equal to the number of bare $\psi^2 \phi^2$ vertices.  Thus provided we have an odd number of $\psi^2 \phi^2$ vertices in the diagram (plus the two external $\psi$ lines), there is no additional rotational invariance constraint and the diagrams are expected to be log divergent.  In other words, if we were to include a $g_4 \Phi^4$ term in the Lagrangian, we would expect the $g_4^2$ contribution to the beta function to vanish and the first nonzero contributions to be order $g_4 g^2$ and $g_4^3$.

\subsection{One Loop Calculations}

The relevant Feynman rules are collected in Appendix B.
We will calculate self-energies for the photon, photinos, and bulk scalars $X$ and $Y$ in general.  
For the boundary degrees of freedom $\psi$ and $\phi$, we will only present detailed self-energy
calculations in the ${\mathcal N}=1$ case since the self-energies of $\psi$ and $\phi$ are irrelevant to the computation of central charges $b_1$ and $b_2$, and we have already discussed why the beta functions should also vanish for ${\mathcal N}=2$ and 4 SUSY.
To incorporate a nonzero $\theta$, simply replace $g \to g \cos(\alpha)$ below. 

\subsubsection*{Photon Self-Energy}

\begin{figure}
\begin{center}
(a) \includegraphics[width=0.31 \textwidth]{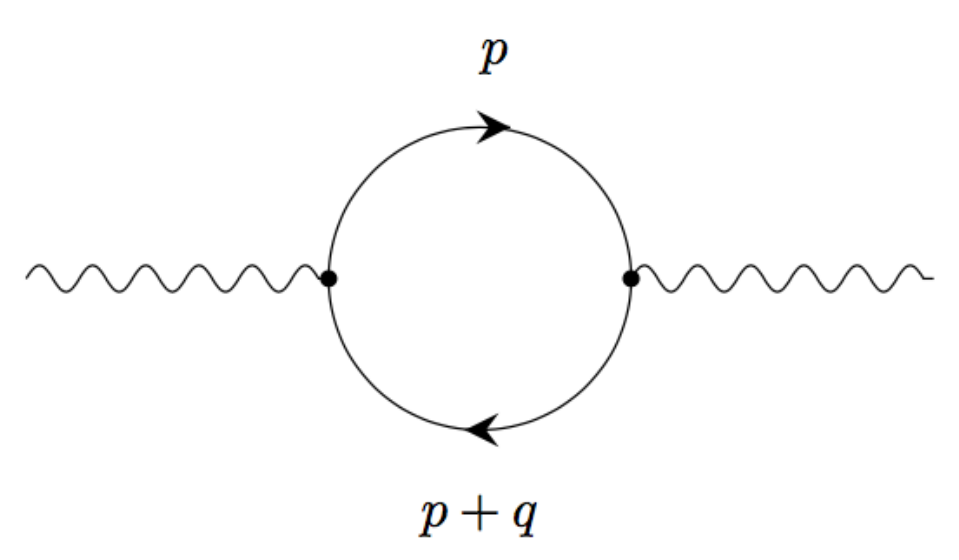}~
(b) \includegraphics[width=0.3 \textwidth]{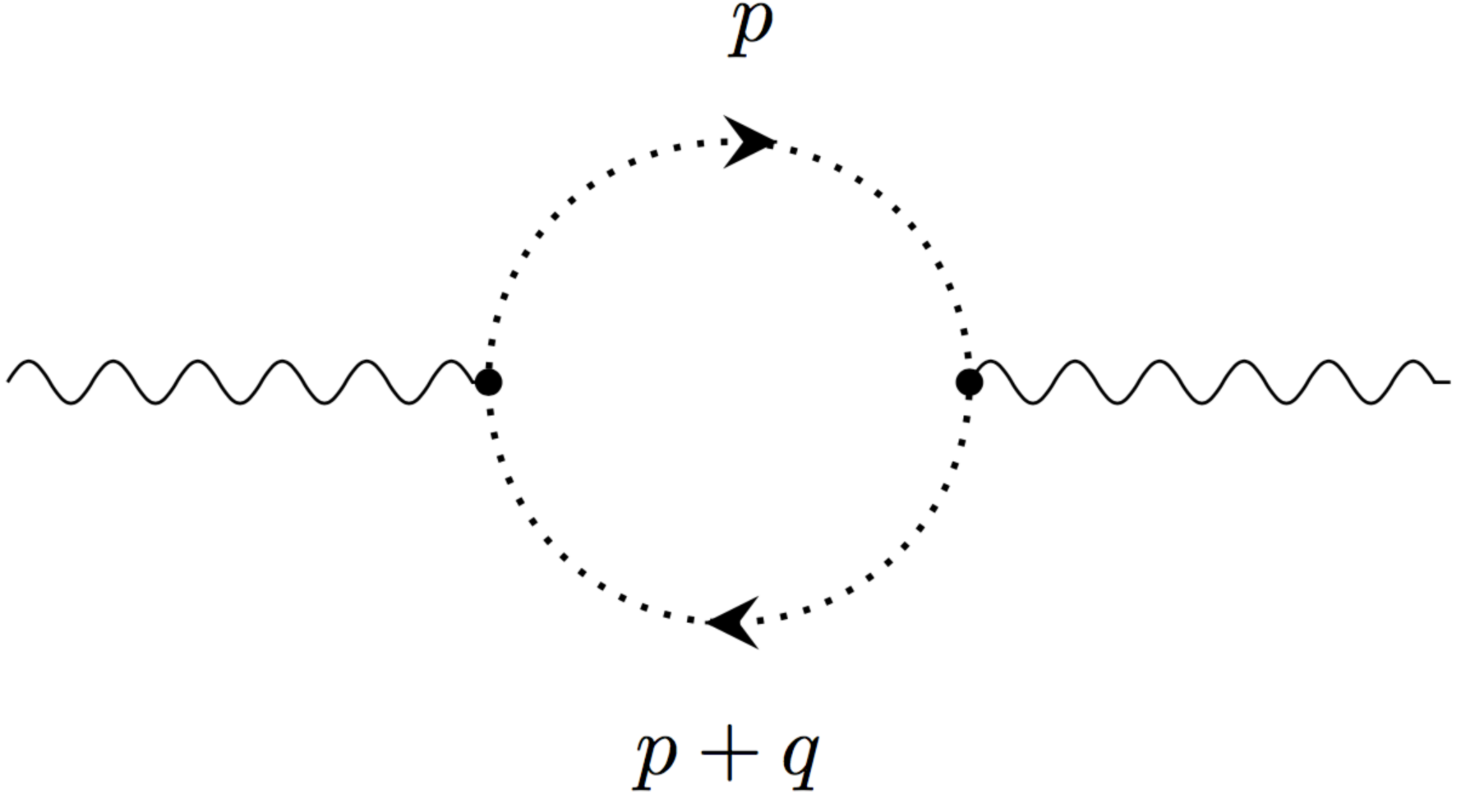}~
(c) \includegraphics[width=0.23 \textwidth]{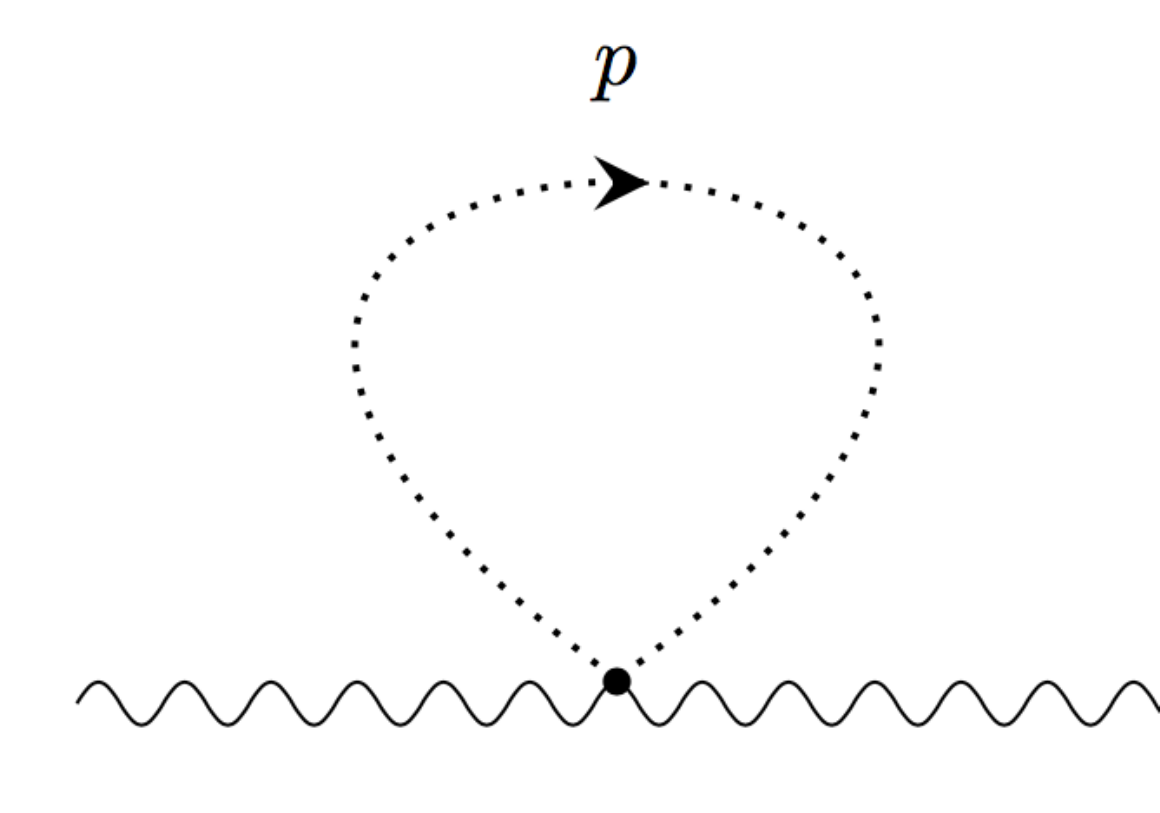}
\end{center}
\caption{The photon self-energy at one-loop.}
\label{fig:photonSE}
\end{figure}

Three diagrams contributed to the photon self-energy at one loop (see figure \ref{fig:photonSE}).
Let us separate out the scalar and fermionic contributions to the photon self-energy:
\be
i \tilde \Pi^{AB}_{(A_\mu)}(q) = i \left(  \tilde \Pi^{AB}_{(A_\mu, \psi)}(q) + \tilde \Pi^{AB}_{(A_\mu, \phi)}(q) \right)  \ .
\ee
In dimensional regularization, we find\footnote{%
We take $\tr \id=2$ for the 3d Clifford space.  Thus, for instance, $\tr (\slashed{p}\slashed{q})=-2 p\cdot q $. 
Note that we adopt effectively 2-component fermions on the boundary so there is a factor 2 difference when comparing \eqref{414} with the corresponding result in \cite{Herzog:2017xha}.
}
\be
\label{414}
i \tilde \Pi^{AB}_{(A_\mu, \psi)} (q)&=&
(-1) (ig)^2 N_f \int \frac{\d^{d-1} p}{(2\pi)^{d-1}} \frac{\tr[\Gamma^A i \slashed{p} \Gamma^B i (\slashed{p} + \slashed{q})]}
{p^2 (p+q)^2} 
\nonumber \\
&=&
 -i g^2 N_f (q^2 \eta^{AB} - q^A q^B)  \frac{(d-3) \pi^{2-\frac{d}{2}}}{4^{d-2} \cos \left( \frac{\pi d}{2} \right) \Gamma \left( \frac{d}{2} \right) } \frac{1}{q^{5-d}} \ , \\
\label{415}
i \tilde \Pi^{AB}_{(A_\mu, \phi)} (q) &=& N_f
\int \frac{\d^{d-1}p}{(2 \pi)^{d-1}} \left( (i g)^2 \left( \frac{1}{i} \right)^2
 \frac{(2p+q)^A (2p+q)^B}{p^2 (p+q)^2}
+(-2i g^2)  \left(\frac{1}{i} \right)\eta^{AB}  \frac{1}{p^2}\right) \nonumber \\
&=& {1\over (d-3)} i \tilde \Pi^{AB}_{(A_\mu, \psi)} (q) \ .
\ee 
We have inserted a factor of $N_f$ to account for the possibility of having $N_f$ flavor multiplets on the boundary.
In $d=4$, the contributions are finite and equal:
\be 
i \tilde \Pi^{AB}_{(A_\mu, \psi)} (q) = i \tilde \Pi^{AB}_{(A_\mu, \phi)} (q) = -\frac{i g^2 N_f }{16 q} (q^2 \eta^{AB} - q^A q^B)\ .
\ee

\subsubsection*{Photino Self-Energy}

\begin{figure}
\begin{center}
\includegraphics[width=0.4 \textwidth]{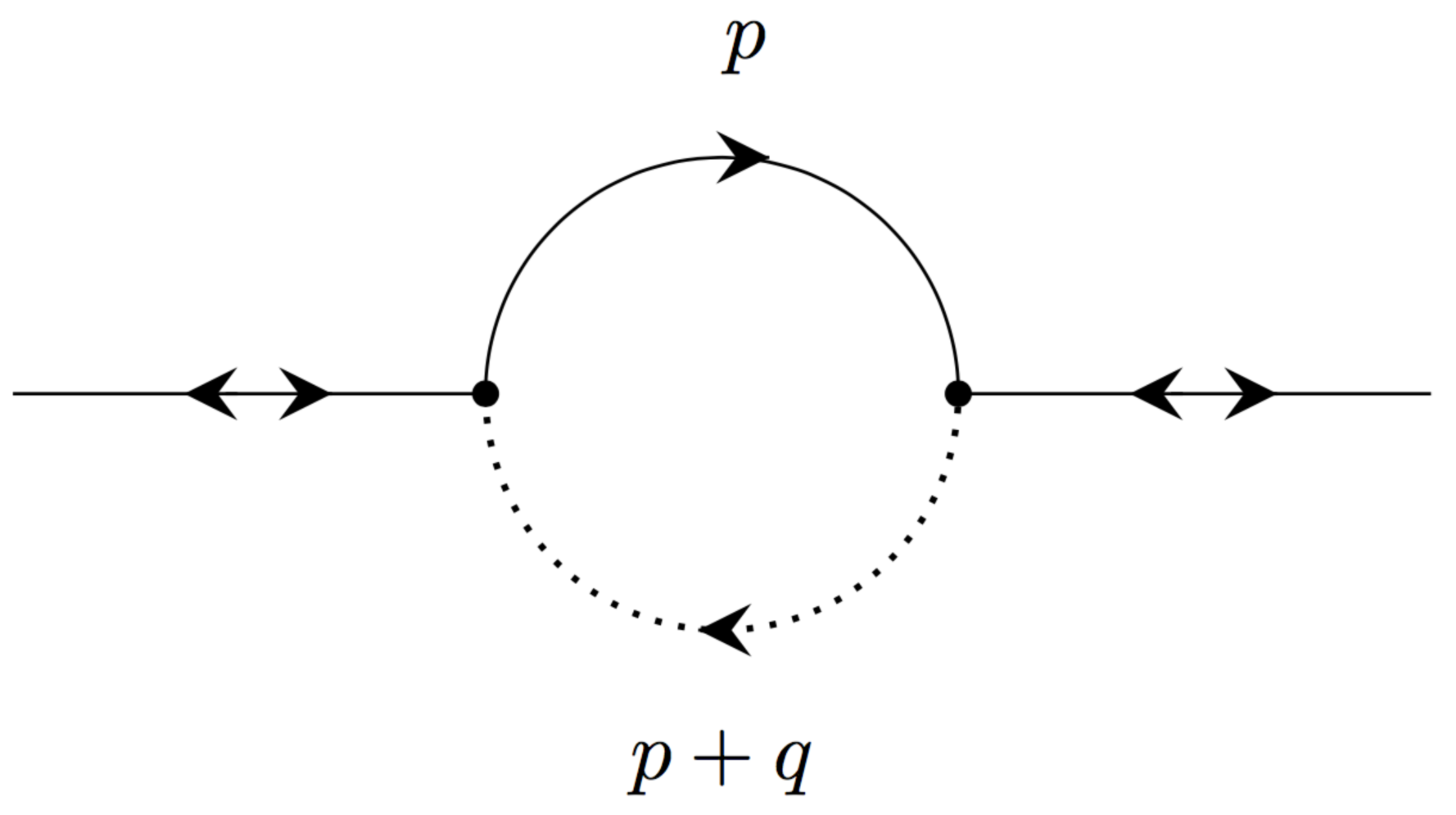}
 \end{center}
 \caption{The photino self-energy at one-loop.}
 \label{fig:photinoSE}
\end{figure}

The photino self-energy at one-loop can be computed from a single diagram (figure \ref{fig:photinoSE}):
\be
i \tilde\Pi_{(\lambda)}(q) 
&=&  2 N_f (g) (-g) \int \frac{\d^{d-1} p}{(2 \pi)^{d-1}} \frac{i \slashed p (-i)}{p^2 (p+q)^2} \nn\\
&=&  \frac{(2  i g^2 N_f)  \pi^{2-\frac{d}{2}} }{4^{d-2}  \cos \left( \frac{\pi d}{2} \right) \Gamma \left( \frac{d}{2} - 1 \right)}
  \frac{\slashed{q}}{q^{5-d} } \ .
  \ee
There is an extra factor of two because there is a second diagram with the charge flowing in the opposite direction inside the loop.  
In 4d, the numerical prefactor of this self-energy is the same as that for the total photon self-energy:
\be
i \tilde\Pi_{(\lambda)}(q) &=& \frac{ i g^2 N_f}{8} \frac{\slashed{q}}{q} \ .
\ee

\subsubsection*{Bulk-Scalars Self-Energy}

\begin{figure}
\begin{center}
(a)\includegraphics[width=0.32\textwidth]{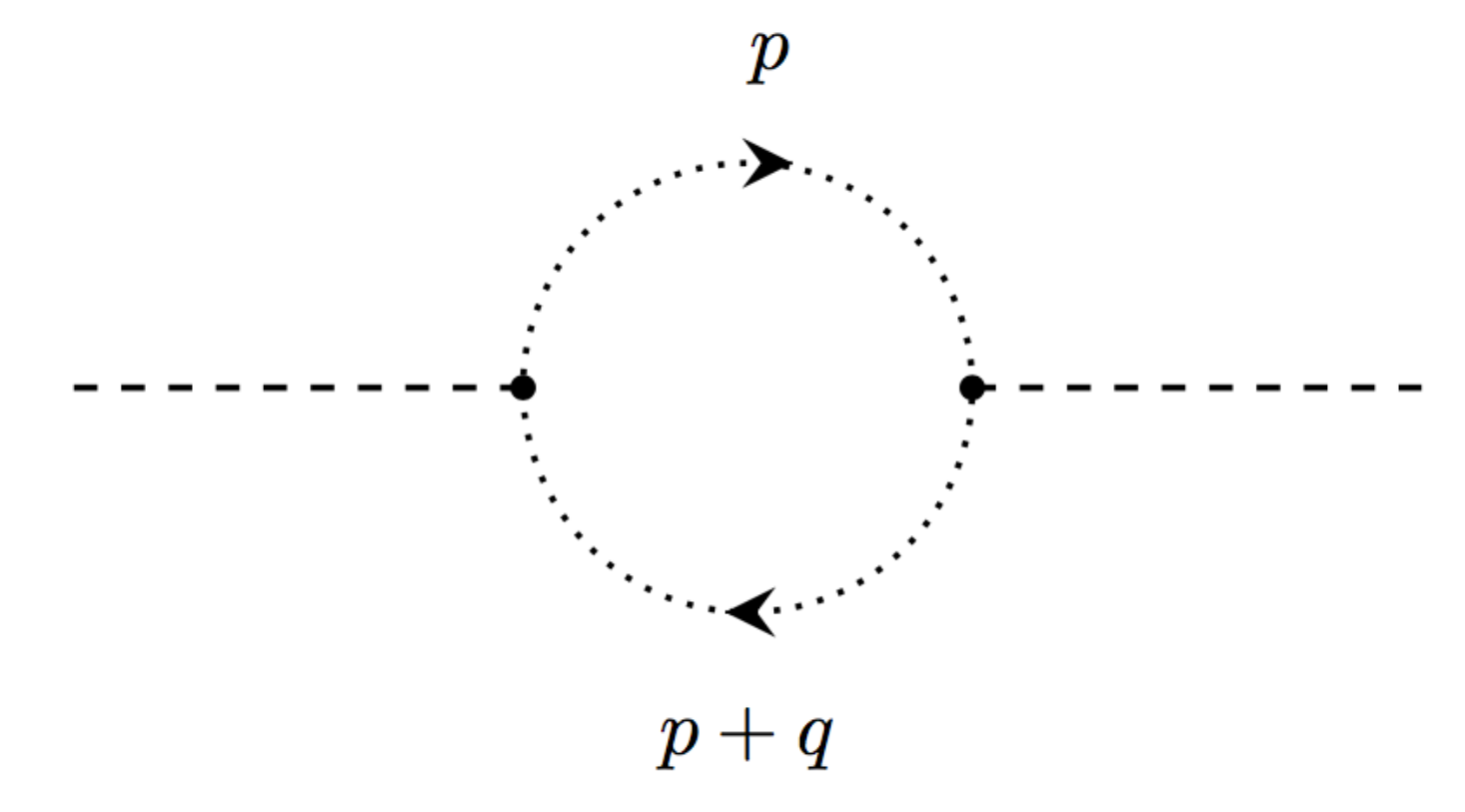}~
(b)\includegraphics[width=0.32 \textwidth]{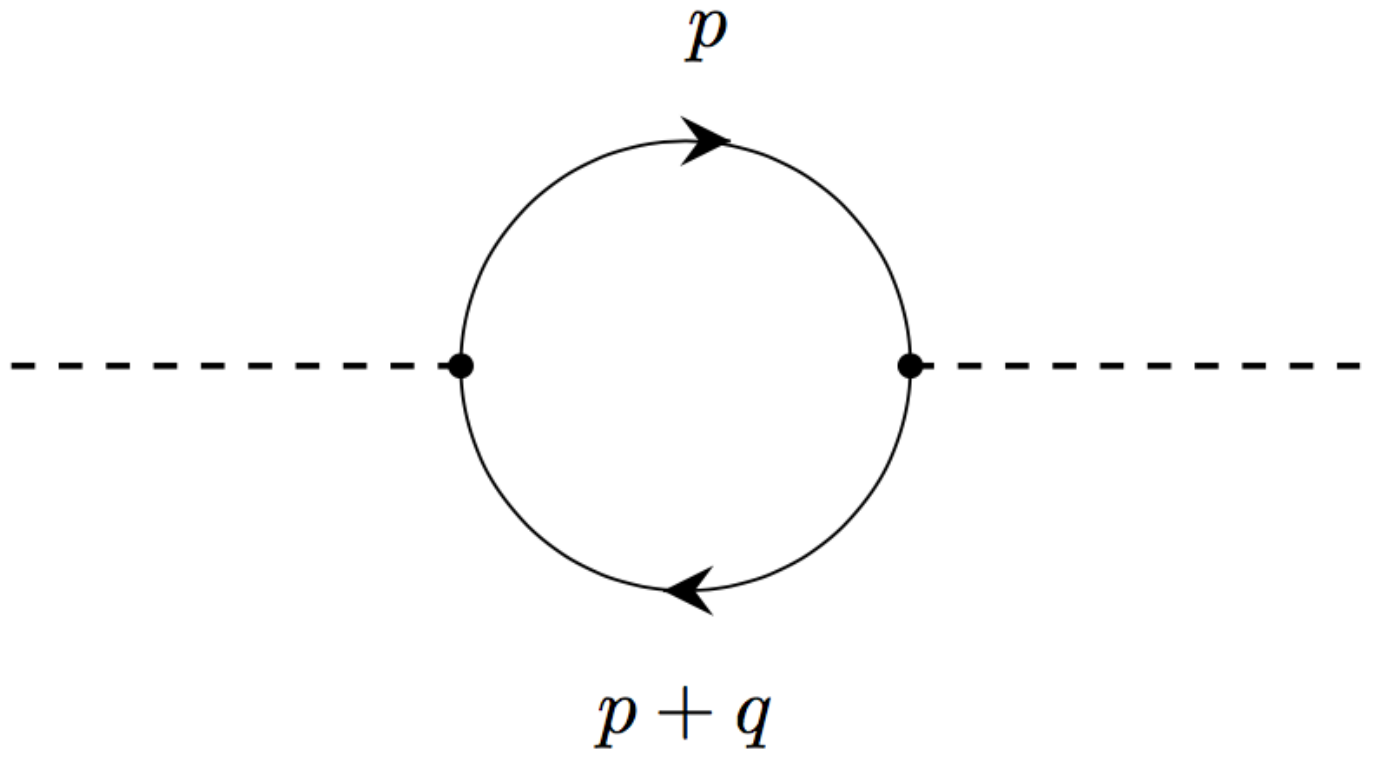}~
(c)\includegraphics[width=0.25 \textwidth]{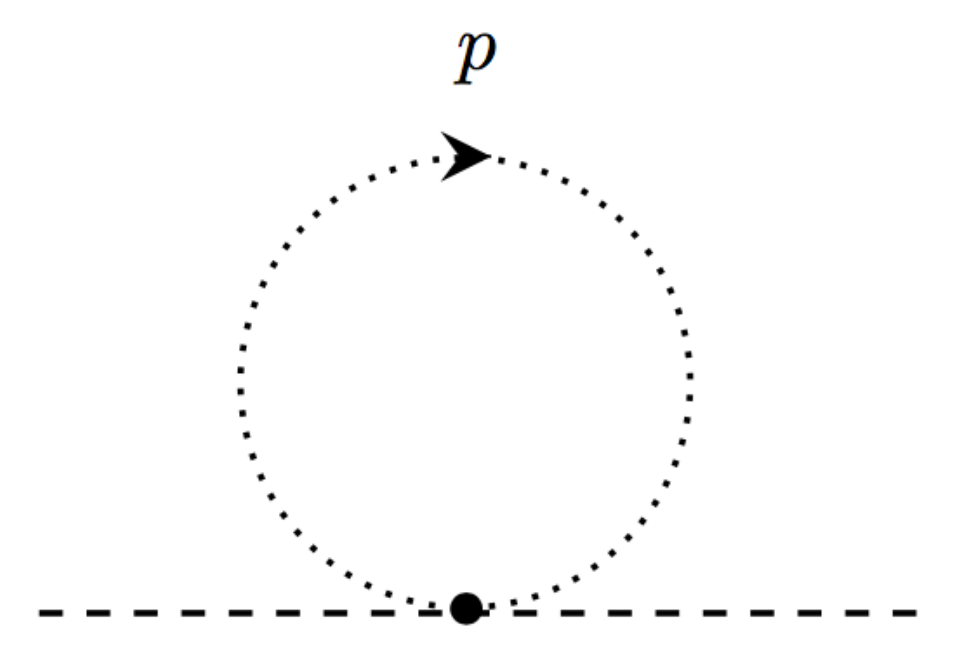}
 \end{center}
 \caption{(a) contributes to the $X$ self-energy; (b), (c) contribute to the $Y$ self-energy. 
 }
 \label{fig:XYSE}
\end{figure}

One diagram contributes to the self-energy of the $X$ scalar (figure \ref{fig:XYSE}a):
\be
\label{423}
i\tilde\Pi_{(X)}(q) &=& (-i g)^2 N_f \int \frac{\d^{d-1} p}{(2\pi)^{d-1}} \frac{q^2 (-i)(-i)}{p^2 (p+q)^2} \nn\\
&=& i g^2 N_f \frac{2^{5-2d} \pi^{2 - \frac{d}{2}} }{\cos \left( \frac{\pi d}{2} \right) \Gamma \left( - 1 + \frac{d}{2} \right)} q^{d-3}\ . 
\ee
Two diagrams contribute to the self-energy of the $Y$ scalar (figures \ref{fig:XYSE}b and \ref{fig:XYSE}c):
\be
\label{422}
i\tilde\Pi_{(Y)}(q)  &=& (-i g)^2 N_f (-1) \int \frac{\d^{d-1} p}{(2\pi)^{d-1}} \frac{\tr[ i \slashed{p} i (\slashed{p}+\slashed{q})]}{p^2 (p+q)^2}
- 2 i g^2 N_f \int \frac{\d^{d-1} p}{(2\pi)^{d-1}} \frac{(-i)}{p^2} \nonumber \\
&=& - i g^2 N_f \frac{2^{5-2d} \pi^{2 - \frac{d}{2}} }{\cos \left( \frac{\pi d}{2} \right) \Gamma \left( -1 + \frac{d}{2} \right) } q^{d-3}\ . 
\ee 

In 4d, the self-energies are equal and opposite: 
\be
i\tilde\Pi_{(X)}(q) =  - i\tilde\Pi_{(Y)}(q) =  \frac{ig^2 N_f}{8}q \ . 
\ee
(The numerical coefficient is the same as in the photon and photino cases.)
Note that here we should not combine \eqref{422} with \eqref{423} as there are two different scalars in the bulk with different boundary conditions.

\subsubsection*{Electron Self-Energy}

\begin{figure}
\begin{center}
(a)\includegraphics[width=0.45 \textwidth]{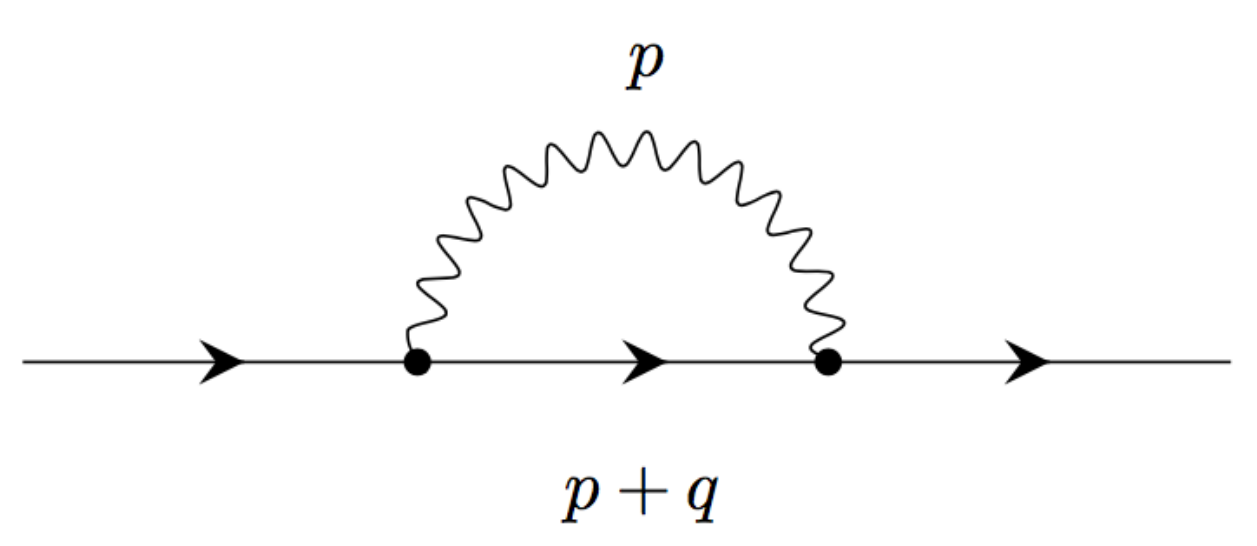}~~~~~
(b)\includegraphics[width=0.35 \textwidth]{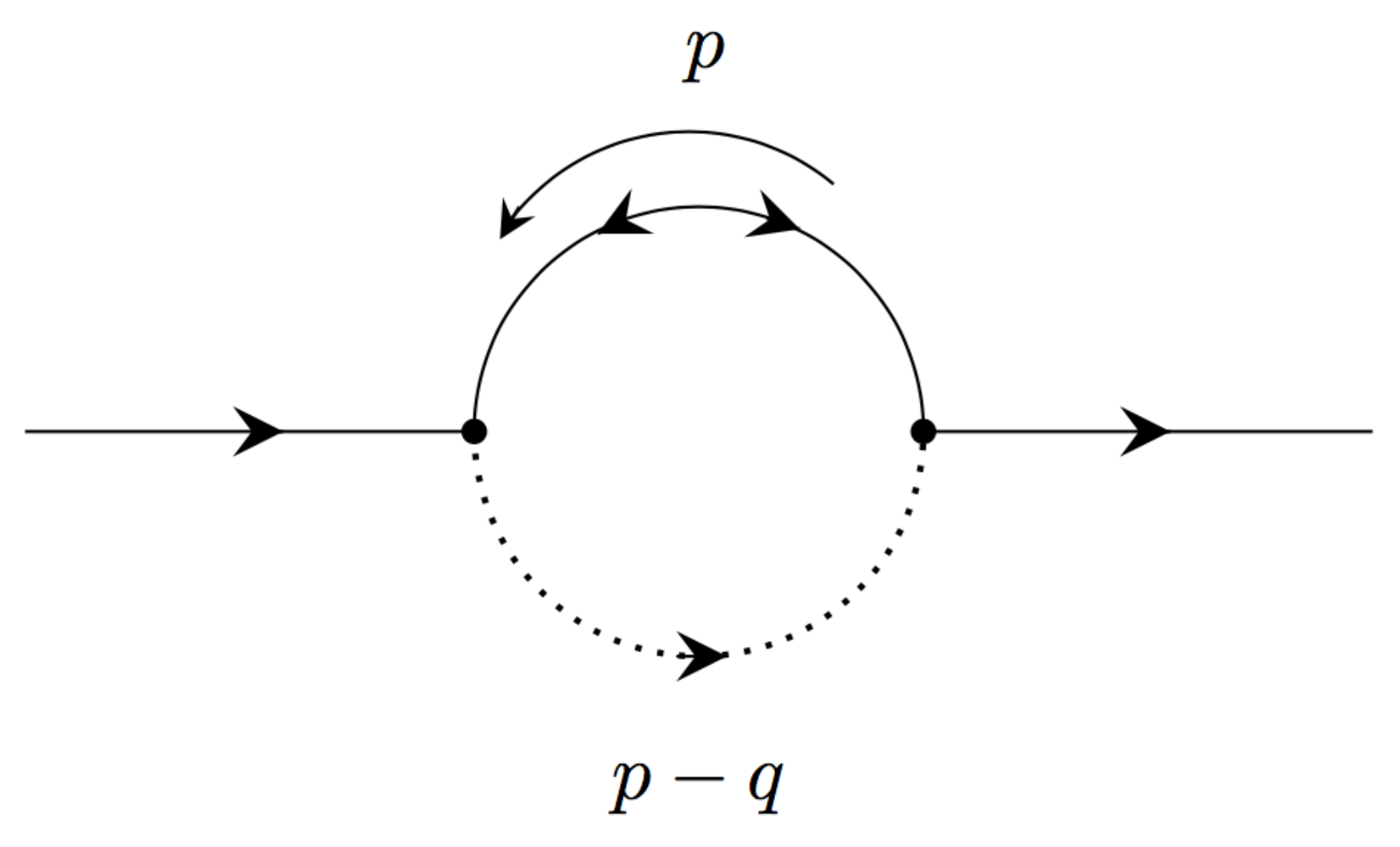}
\end{center}
\caption{The electron self-energy at one-loop in the ${\mathcal N}=1$ theory. 
\label{fig:electronSE}}
\end{figure}

There are both photon and photino contributions to the electron self-energy
(figure \ref{fig:electronSE}):
\be
i \tilde \Pi_{(\psi)}(q) = i \left(\tilde \Pi_{(\psi, A_\mu)}(q) + \tilde \Pi_{(\psi, \lambda)}(q)\right) \ ,
\ee
where 
\be
i \tilde \Pi_{(\psi, A_\mu)}(q) &=& (ig)^2 \int \frac{\d^{d-1} p}{(2\pi)^{d-1}} \frac{ \Gamma^A i \slashed{p} \Gamma^B (-i) \eta_{AB}}{p^2 |p-q|}
= - \frac{ig^2}{6\pi^2 \epsilon} \slashed{q} + \ldots \ , \\
i \tilde \Pi_{(\psi, \lambda)}(q) &=&  (g) (- g) \int \frac{\d^{d-1}p}{(2\pi)^{d-1}} \frac{ i\slashed{p} }{|p|} \frac{(-i)}{ (p-q)^2} =  -\frac{ i g^2}{3 \pi^2 \epsilon} \slashed{q}+ \ldots \ .
\ee 
We focus on the singular contribution to the diagram in $d=4-\epsilon$ dimensions.  
The total  singular  contribution is
\be
i \tilde \Pi_{(\psi)}(q) = - \frac{i g^2}{2 \pi^2 \epsilon} + \ldots \ .
\ee

\subsubsection*{Selectron Self-Energy}

\begin{figure}
\begin{center}
(a)\includegraphics[width=0.33 \textwidth]{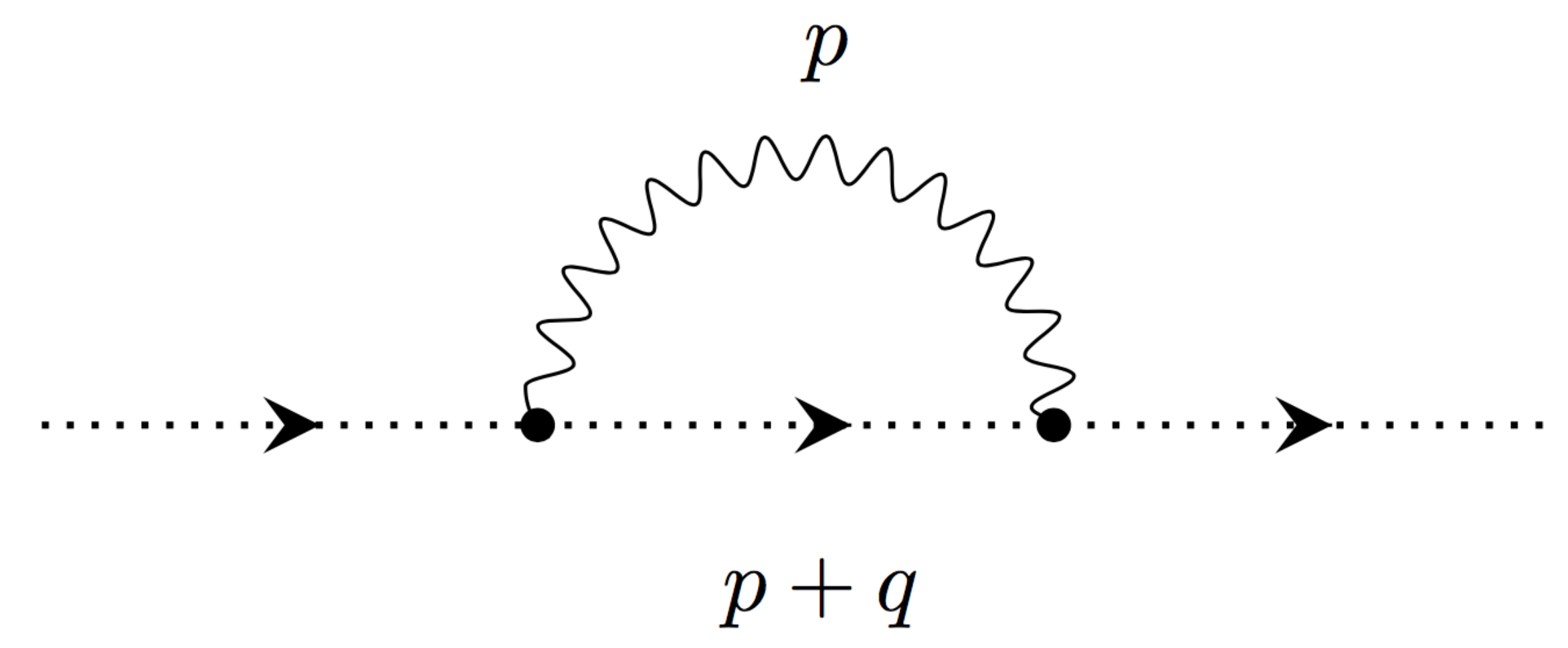}
(b)\includegraphics[width=0.24 \textwidth]{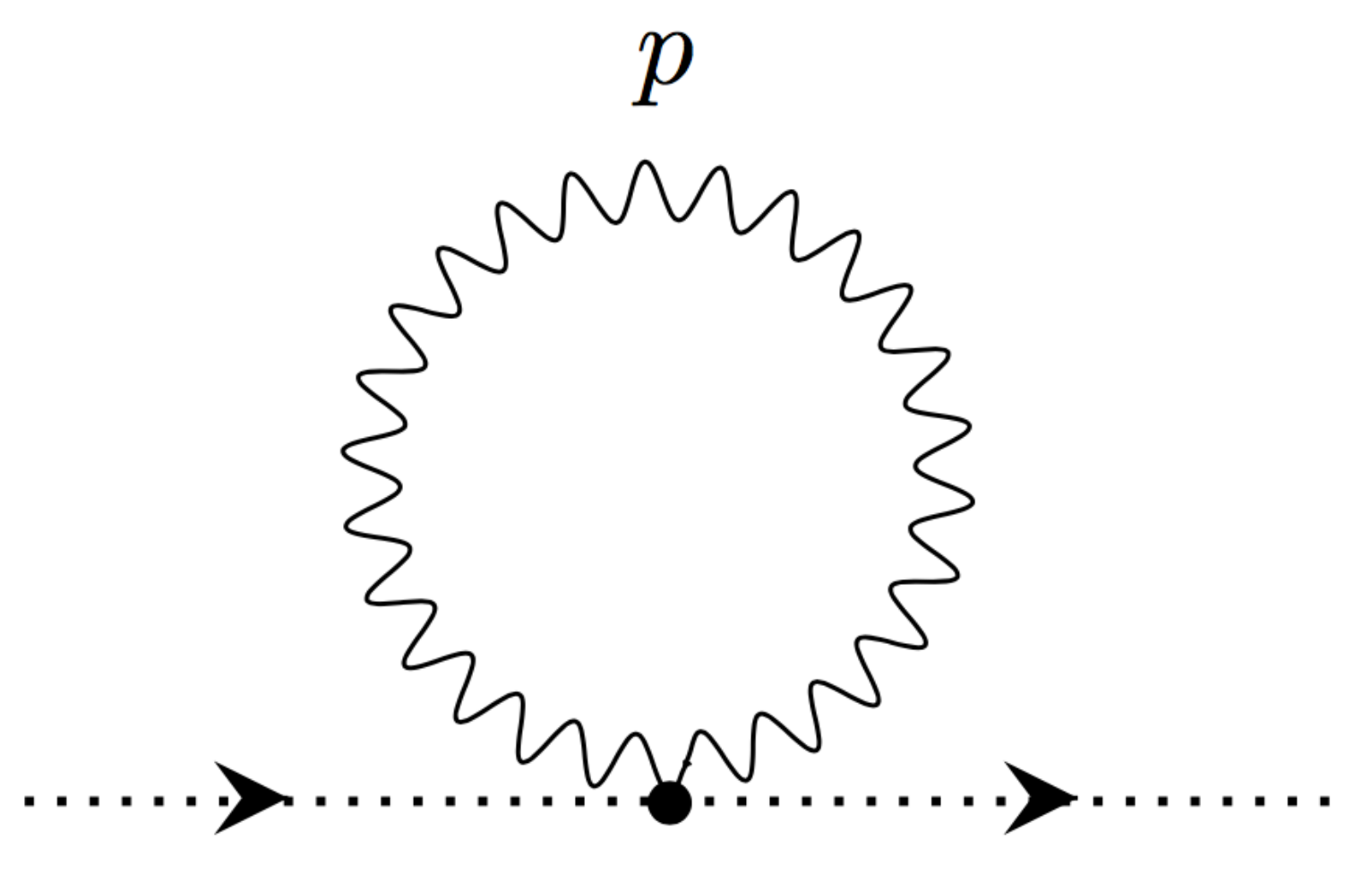}
(c)\includegraphics[width=0.31\textwidth]{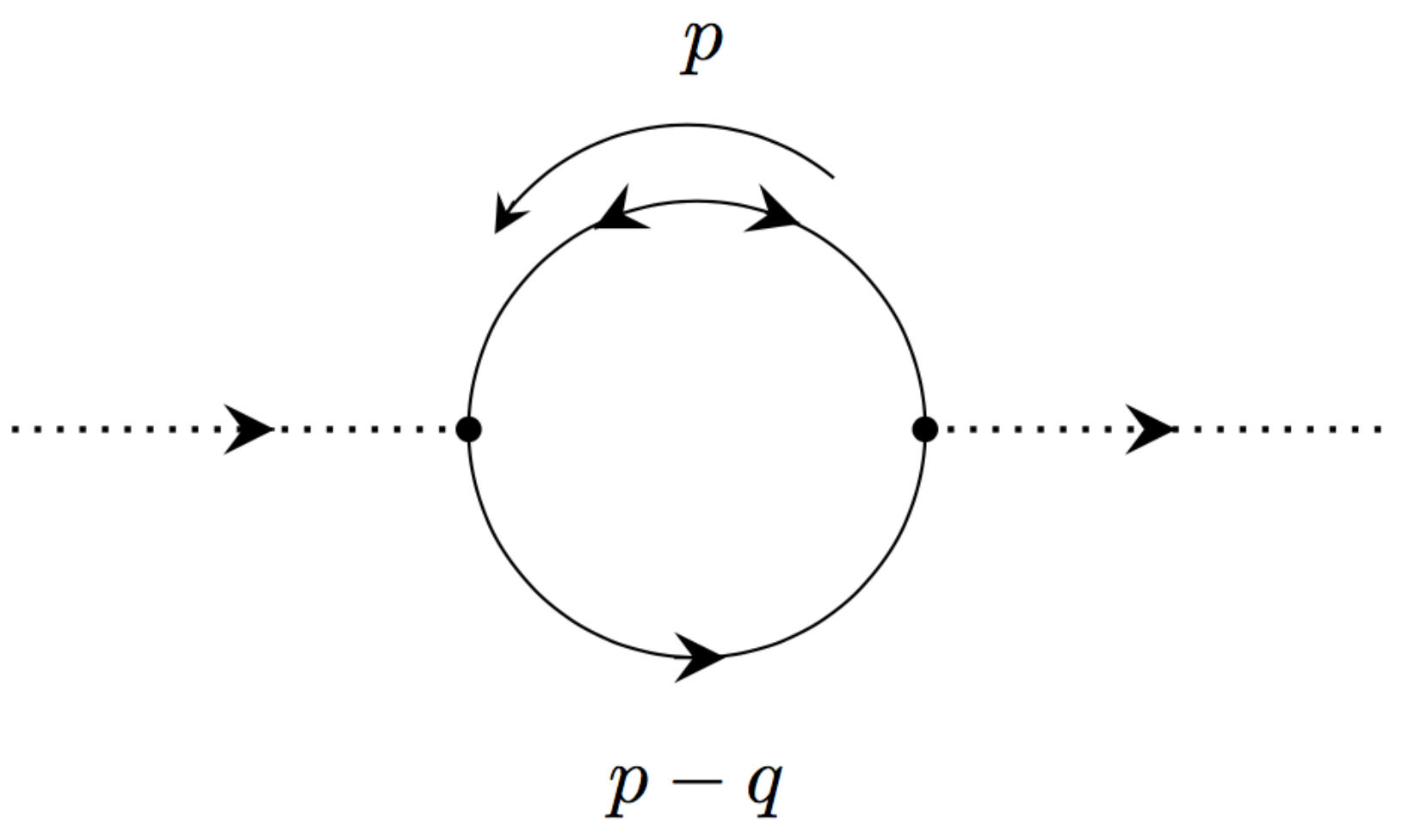}
\end{center}
\caption{The selectron self-energy at one-loop in the ${\mathcal N}=1$ theory. 
\label{fig:selectronSE}}
\end{figure}

We again divide up the one-loop self-energy into contributions from photons and contributions with photinos running in the loop
(figure \ref{fig:selectronSE}):
\be
i \tilde \Pi_{(\phi)}(q^2) = i \left(\tilde \Pi_{(\phi, A_\mu)}(q^2) + \tilde \Pi_{(\phi, \lambda)}(q^2) \right) \ ,
\ee
where 
\be
i \tilde \Pi_{(\phi, A_\mu)}(q^2)  &=& (i g)^2 \left(-i \right)^2 \int \frac{\d^{d-1} p}{(2\pi)^{d-1}}
\frac{(p+q)^2}{|p-q| p^2} + (-2i g^2) (d-1) \left(-i \right) \int \frac{\d^{d-1} p}{(2\pi)^{d-1}} \frac{1}{|p|} \nonumber \\
&=& \frac{5 i g^2}{6 \pi^2 \epsilon} q^2 + \ldots\ , \\
i \tilde \Pi_{(\phi, \lambda)}(q^2) &=&  (g) (-g) (-1) \int \frac{\d^{d-1} p}{(2\pi)^{d-1}} 
\frac{\tr[ i\slashed{p} i(\slashed{p} - \slashed{q})]}{|p| (p-q)^2} = -\frac{ i g^2}{3 \pi^2 \epsilon} q^2 + \ldots \ .
\ee
The photino contribution comes from a single diagram while
the photon contribution comes from a pair of diagrams.
We again focus on the singular contribution to the diagrams in $d=4-\epsilon$ dimensions.
The total  singular  contribution is 
\be
i \tilde \Pi_{(\phi)}(q^2) = \frac{ig^2}{2 \pi^2 \epsilon} q^2 + \ldots \ .
\ee

\subsubsection*{Charge Renormalization for Electron}

\begin{figure}
\begin{center}
(a) \includegraphics[width=0.44 \textwidth]{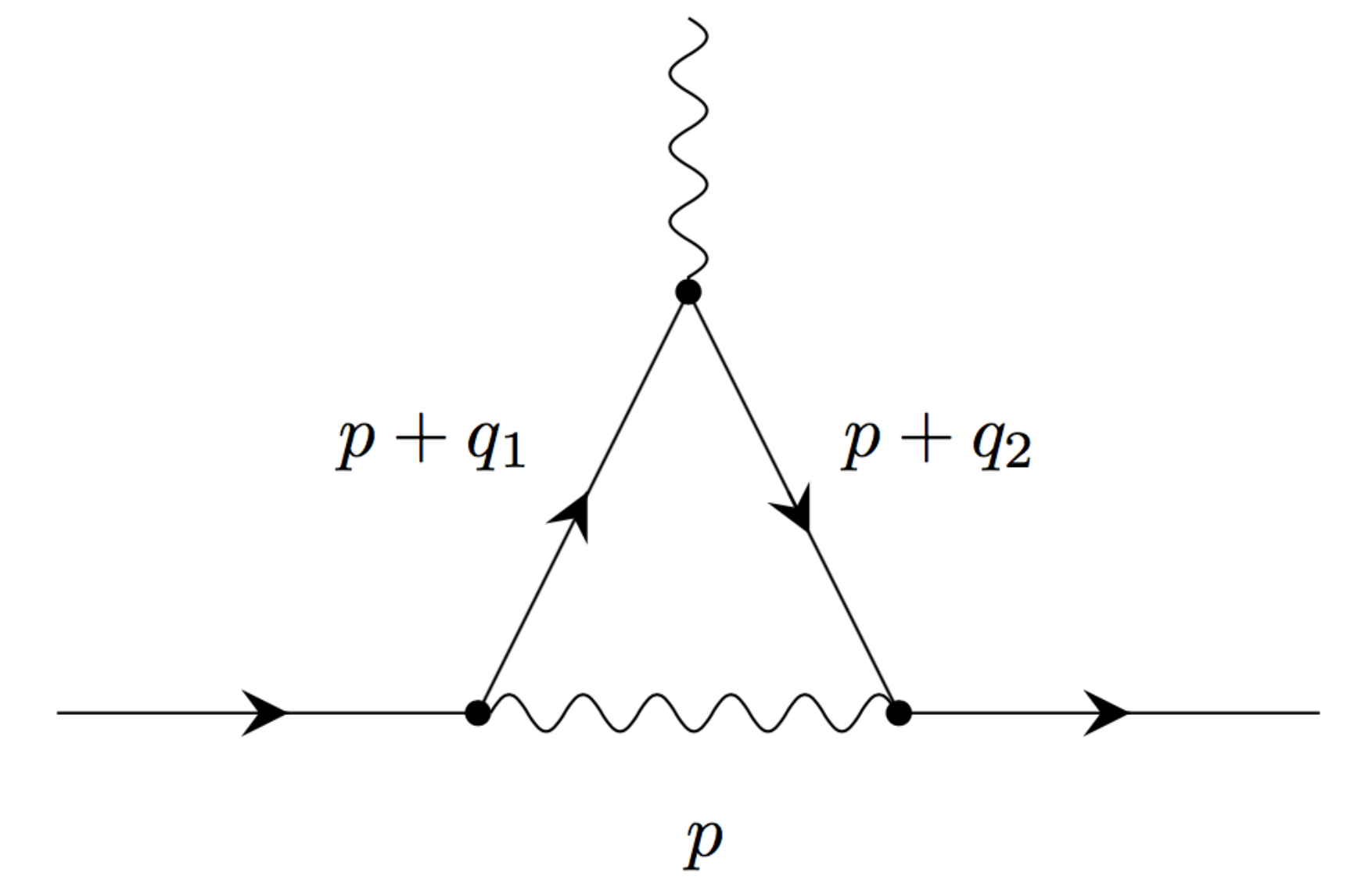}~~~
(b) \includegraphics[width=0.44 \textwidth]{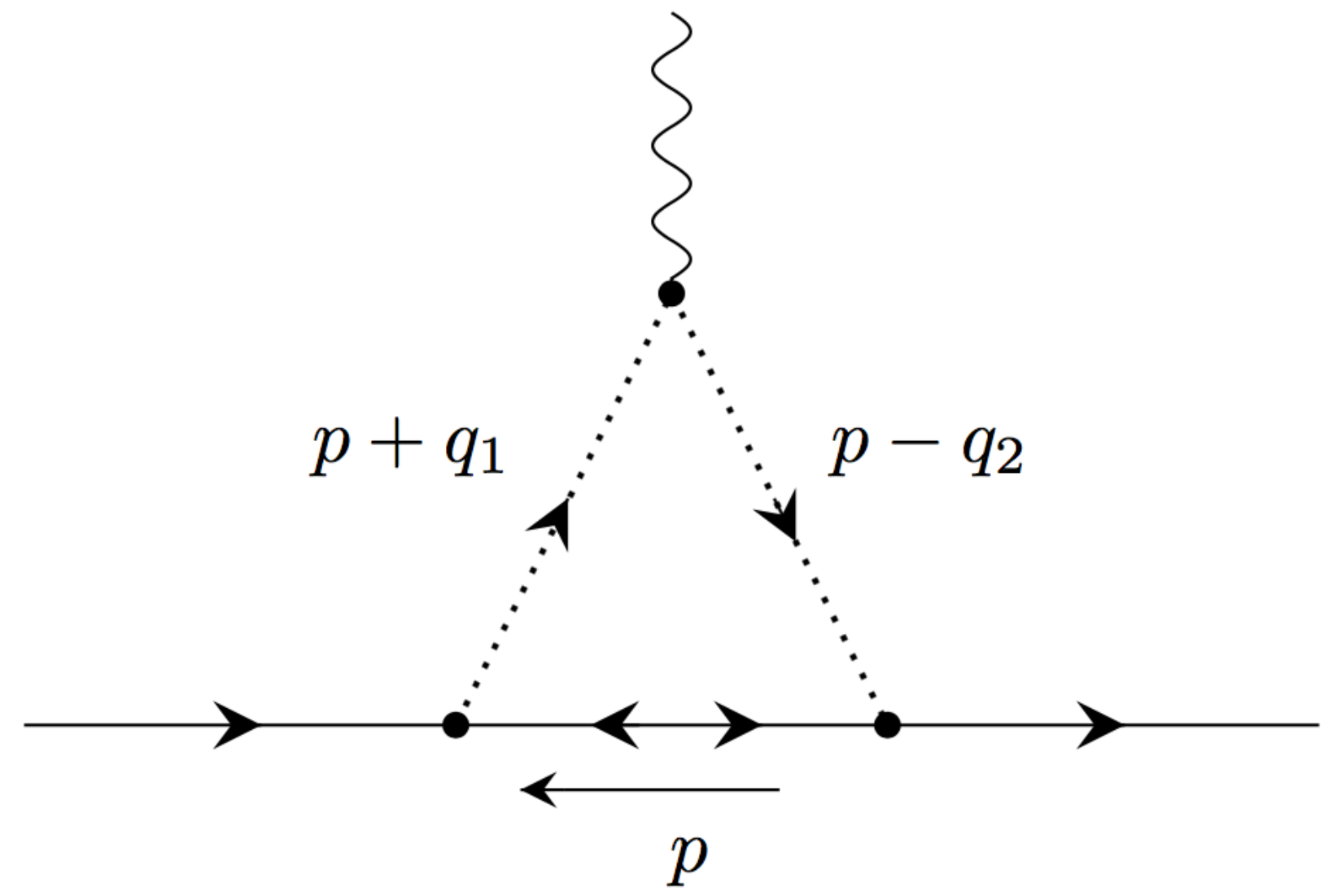}
\end{center}
\caption{The electron charge renormalization in the ${\mathcal N}=1$ theory.
\label{fig:electronvertex}}
\end{figure}

We divide up the vertex renormalization of the electron into contributions with a photon and with a photino running in the loop
(figure \ref{fig:electronvertex}):
\be
i \tilde V^A_{(\psi)} (q_1, q_2) = i \left(\tilde V^A_{(\psi, A_\mu)}(q_1,q_2) + \tilde V^A_{(\psi, \lambda)} (q_1,q_2) \right) \ , 
\ee
where 
\be
i \tilde V^A_{(\psi,A_\mu)} (q_1, q_2) &=& (ig)^3 \int \frac{\d^{d-1} p}{(2\pi)^{d-1}} 
\frac{\Gamma^C i (\slashed{p} + \slashed{q}_1)
\Gamma^A i (\slashed{p} + \slashed{q}_2) \Gamma^B (-i) \eta_{CB}}
{(p+q_1)^2  (p+q_2)^2 |p|} \nn\\
&=& \frac{i g^3 \Gamma^A}{6 \pi^2 \epsilon}+ \ldots \ ,\\
\label{VA2}
i \tilde V^A_{(\psi, \lambda)}(q_1, q_2) &=& (g)(-g)(ig) \int \frac{\d^{d-1} p}{(2 \pi)^{d-1}} \frac{(2 p + q_1 - q_2)^A i(-\slashed{p}) (-i)^2}{|p| (p+q_1)^2 (p-q_2)^2} \nn\\&=&\frac{i g^3 \Gamma^A}{3 \pi^2 \epsilon} + \ldots  
\ .
\ee 
There is one diagram with a photon in the loop and one diagram with a photino.
The total singular contribution is 
\be
i \tilde V_{(\psi)}^A(q_1, q_2) = \frac{i g^3 \Gamma^A}{2 \pi^2 \epsilon} + \ldots \ .
\ee

\subsubsection*{Charge Renormalization for Selectron}

\begin{figure}
\begin{center}
(c)\includegraphics[width=0.44 \textwidth]{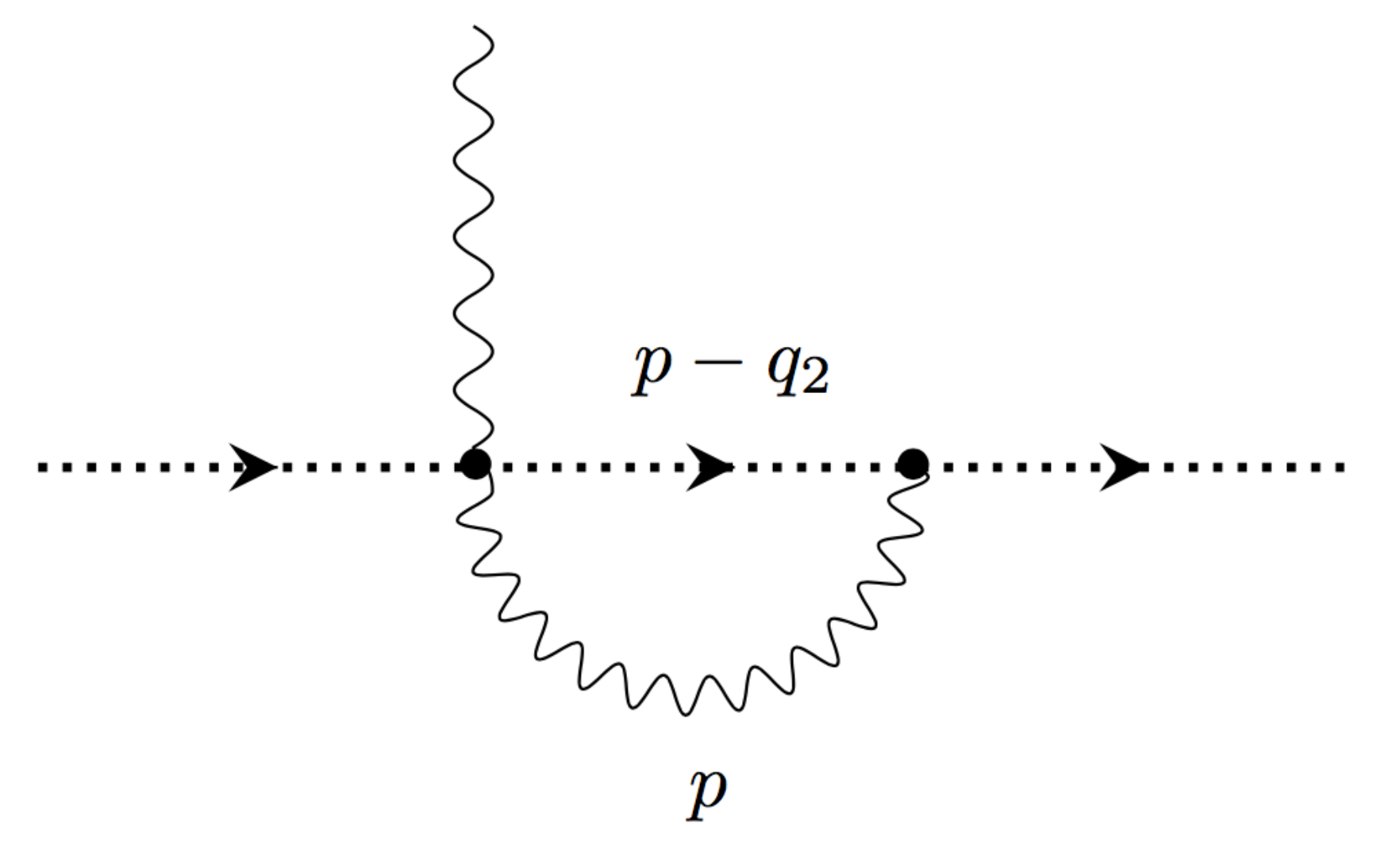}
(d)\includegraphics[width=0.44 \textwidth]{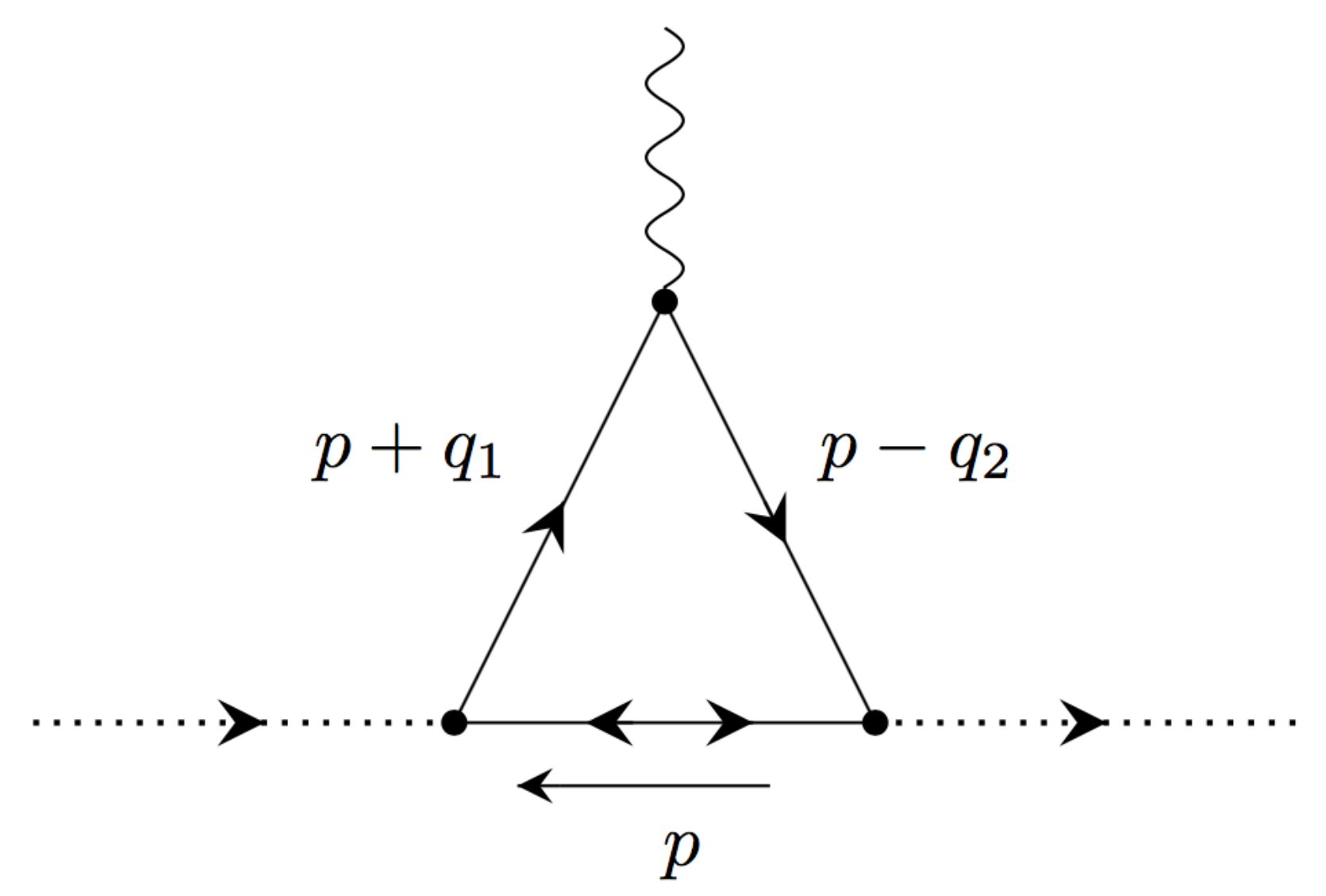}
\end{center}
\caption{The selectron charge renormalization in the ${\mathcal N}=1$ theory.
\label{fig:selectronvertex}}
\end{figure}

We divide up the vertex renormalization into contributions with photons and photinos
(figure \ref{fig:selectronvertex}):
\be
i \tilde V^A_{(\phi)} = i \left(\tilde V^A_{(\phi, A_\mu)} + \tilde V^A_{(\phi, \lambda)} \right) \ .
\ee
There are three diagrams  with a photon running in the loop:
\be
i \tilde V_{(\phi, A_\mu)}^A (q_1, q_2) &=& (i g)^3 \int \frac{ \d^{d-1} p}{(2 \pi)^{d-1}} \frac{(p + 2 q_1) \cdot (p - 2 q_2) (2p +q_1 - q_2)^A
(-i)^3}
{(p+q_1)^2 (p-q_2)^2 |p|}  \nonumber \\
&& + 
(-2 i g^2) (i g) \int \frac{ \d^{d-1} p}{(2 \pi)^{d-1}} \left(  \frac{(p + 2 q_1)^A (-i)^2}{(p+q_1)^2 |p|}+ 
\frac{(p - 2 q_2)^A (-i)^2}{(p-q_2)^2 |p|}  \right) \nonumber \\
&=& -i g^3 \frac{1}{2 \pi^2 \epsilon} (q_1 - q_2)^A \left(-1 +\frac{8}{3}\right)+ \ldots
\ee
The flip in sign of $q_2$ with respect to the Feynman rule is because we take $q_2$ to be ingoing.
There is only one diagram with a photino in the loop:
\be
i \tilde V_{(\phi, \lambda)}^A (q_1, q_2) &=&
 (g) (-g) (i g) (-1) \int \frac{\d^{d-1} p}{(2 \pi)^{d-1}}
\frac{\tr[i(\slashed{p} + \slashed{q}_1) \Gamma^A i(\slashed{p} - \slashed{q}_2) i\slashed{p}]}
{(p+q_1)^2 (p-q_2)^2 |p| } \nonumber \\
&=&  i\frac{ g^3}{3 \pi^2 \epsilon} (q_1 - q_2)^A + \ldots
\ee 
The total singular contribution is 
\be
i \tilde V^A_{(\phi)} = - \frac{ i g^3}{2 \pi^2 \epsilon}(q_1 - q_2)^A + \ldots \ .
\ee

\begin{figure}
\begin{center}
(a)\includegraphics[width=0.44 \textwidth]{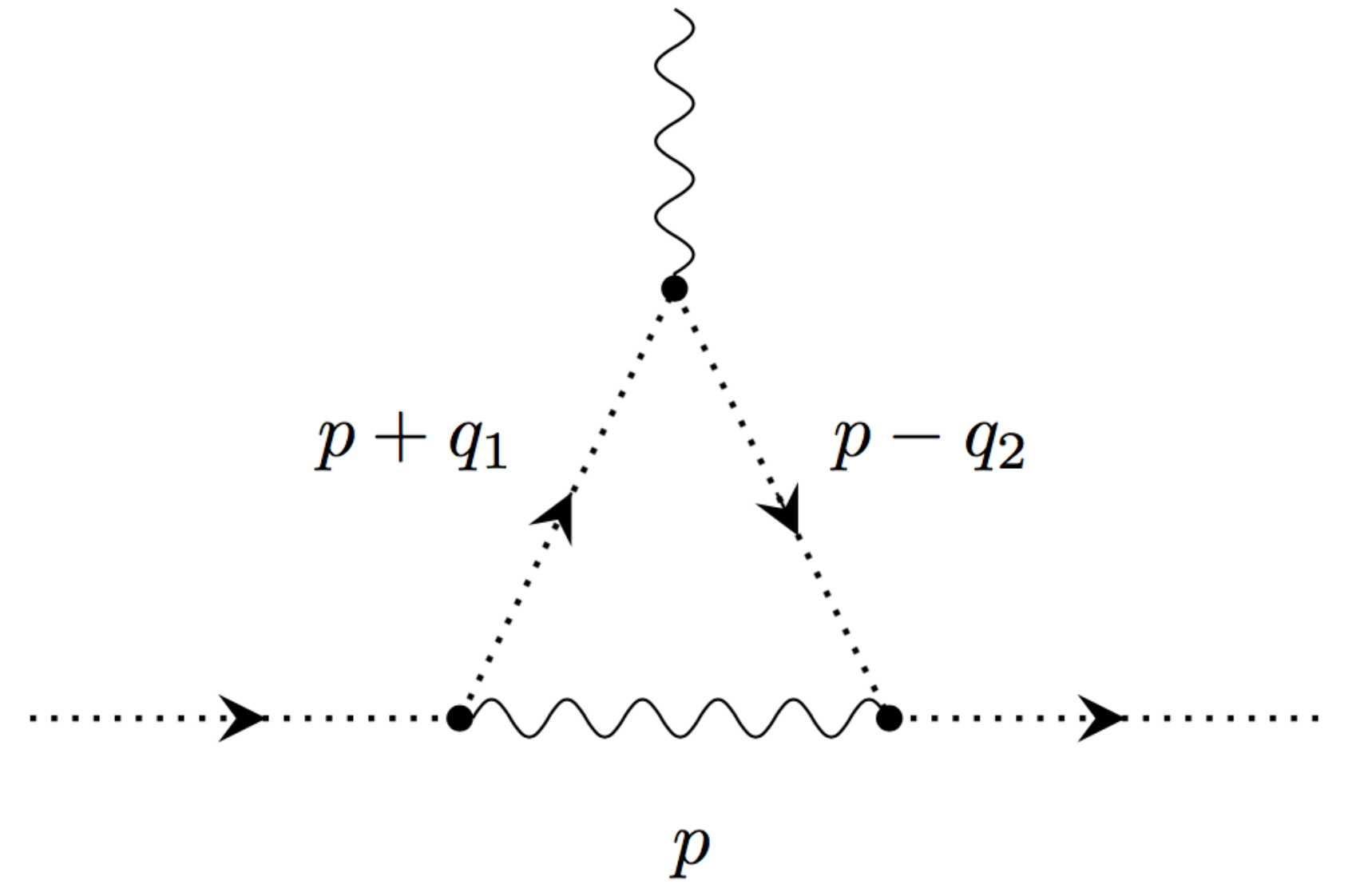}
(b)\includegraphics[width=0.44 \textwidth]{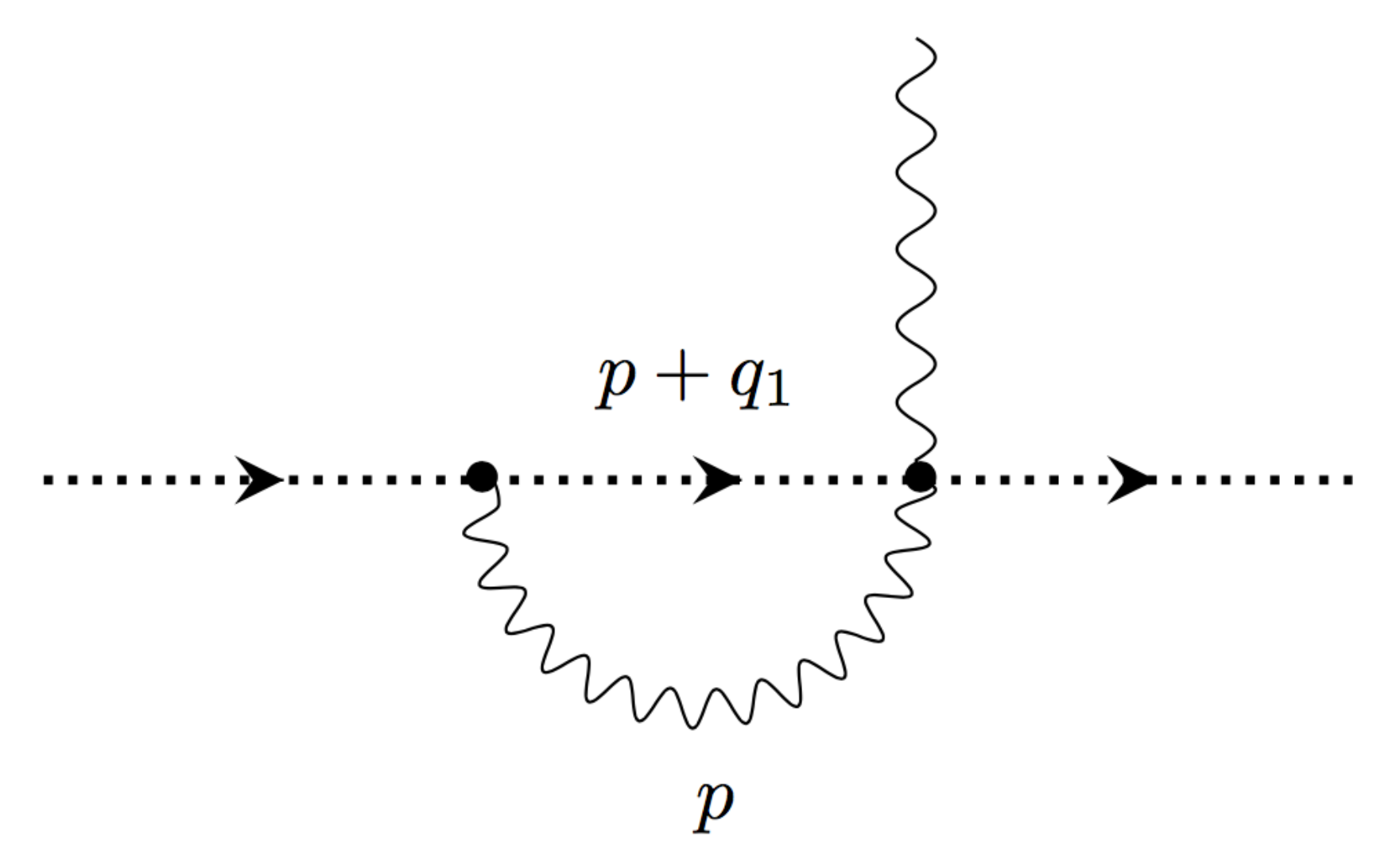}
\end{center}
\end{figure}

\subsubsection*{Yukawa Renormalization}

\begin{figure}
\begin{center}
(a) \includegraphics[width=0.44 \textwidth]{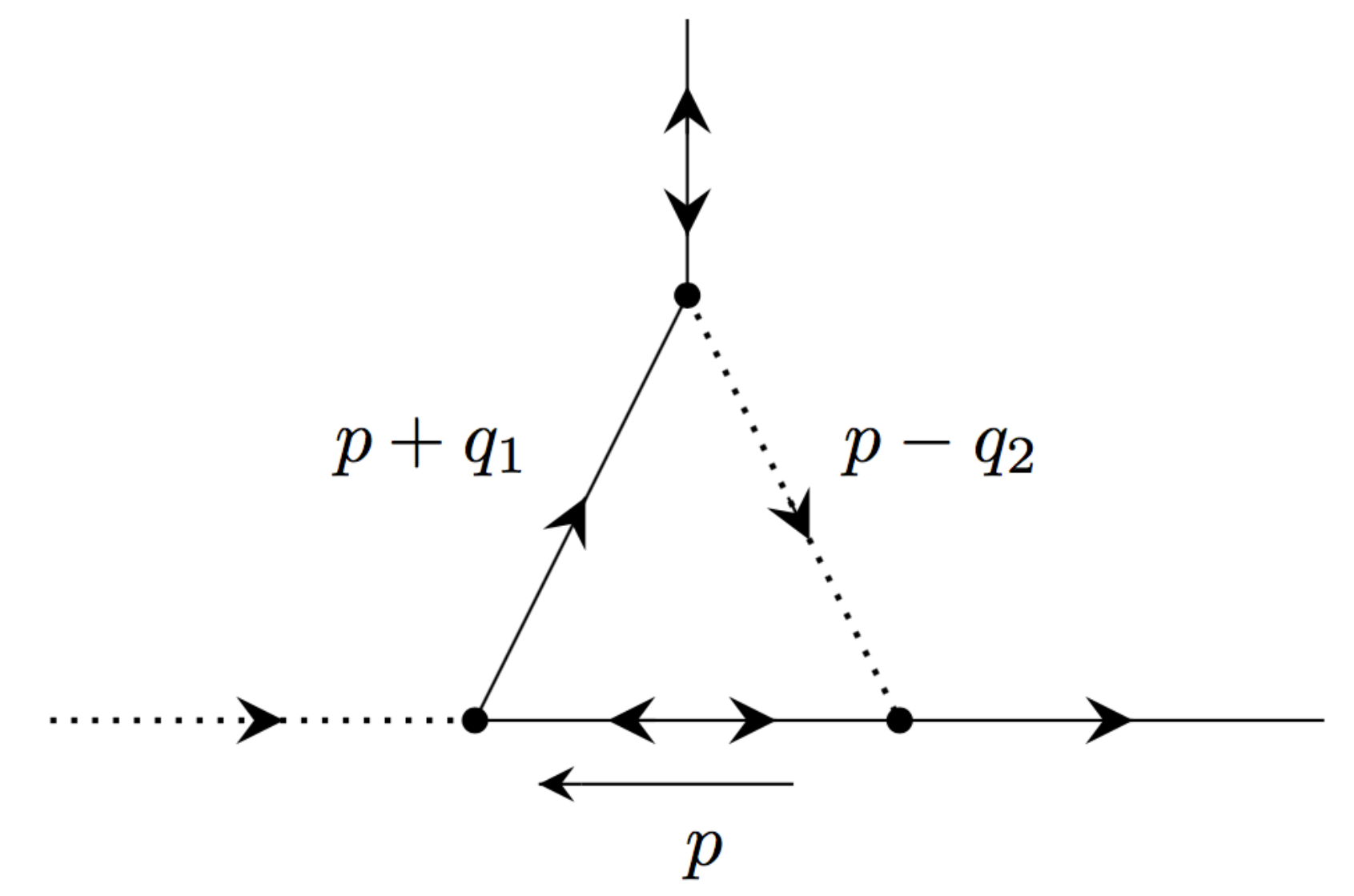}~~
(b) \includegraphics[width=0.44 \textwidth]{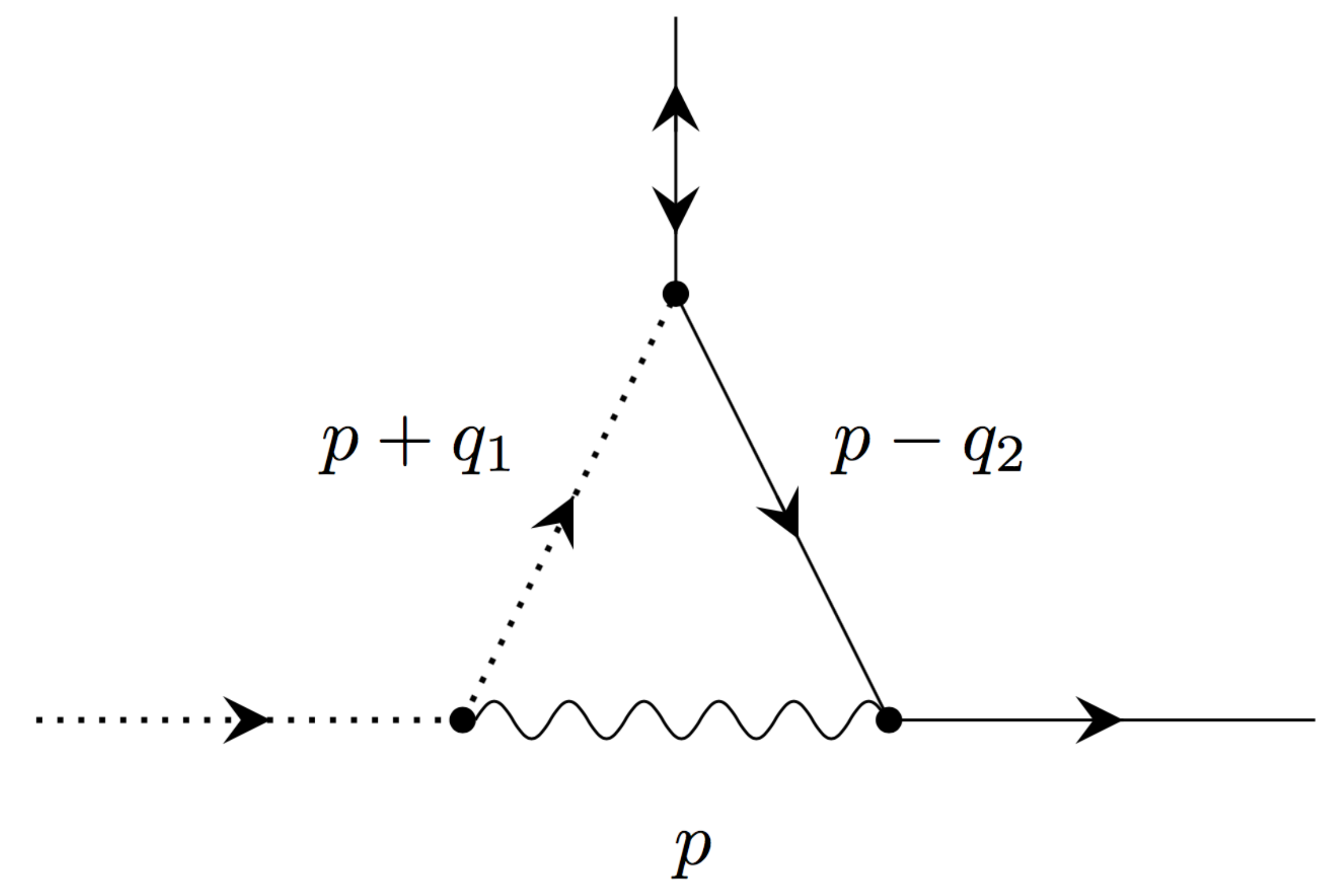}
\end{center}
\caption{The Yukawa coupling renormalization in the ${\mathcal N}=1$ theory.
\label{fig:Yukawavertex}
}
\end{figure}

Two diagrams contribute at one-loop to the renormalization of the Yukawa $\phi \psi \lambda$ interaction
(figure \ref{fig:Yukawavertex}):
\be
i \tilde V_{(Y)} = i \left( \tilde V_{(Y,\lambda)} + \tilde V_{(Y, A_\mu)} \right) \ ,
\ee
where
\be
i \tilde V_{(Y, \lambda)}(q_1, q_2) &=&   (g)^2 (-g) \int \frac{\d^{d-1} p}{(2 \pi)^{d-1}} \frac{ i\slashed{p}\, i (\slashed{p} + \slashed{q}_1)  (-i)}
{|p| (p+q_1)^2  (p-q_2)^2}
=  - \frac{g^3}{2 \pi^2} \frac{1}{\epsilon} + \ldots \ ,\\
i \tilde V_{(Y, A_\mu)} &=& (g) (ig)^2 \int \frac{\d^{d-1} p}{(2\pi)^{d-1}}
\frac{(-i) (-i) i(\slashed{p} - \slashed{q}_2) (\slashed{p} + 2\slashed{q}_1) }{(p+q_1)^2 |p| (p-q_2)^2} 
=   \frac{g^3}{2\pi^2} \frac{1}{\epsilon} + \ldots
\ee
The total singular contribution vanishes.

\section{Conclusion and 
Future Perspective}

We have demonstrated that ${\mathcal N}=1$, ${\mathcal N}=2$, and ${\mathcal N}=4$ super graphene models are all very likely to be interesting examples of boundary superconformal field theories. Their beta functions vanish at all orders in perturbation theory in $d=4$.  Hence, the gauge coupling $g$ should be exactly marginal.  We have shown how the boundary central charges $b_1$ and $b_2$ depend explicitly on $g$.  
Unlike the situation for the bulk central charge $c$, neither $b_1$ or $b_2$ is protected by supersymmetry.\footnote{%
Note the fact that these boundary central charges depend on marginal couplings naively suggests that they are unlikely to be useful candidates to measure boundary renormalization group flow in 4d bCFTs.  
In this sense, these boundary charges are rather different from the $a$-central charge in 4d, the $c$-central charge in 2d, and the 3d sphere partition function -- often called $f$ -- which are known to be larger in the UV than in the IR and are independent of marginal couplings \cite{Komargodski:2011vj,Gerchkovitz:2014gta,Zamolodchikov:1986gt}.
}

There is however one interesting caveat for ${\mathcal N}=4$ theories, where we noticed that the combination $b_1^{({\mathcal N}=4)} - b_2^{({\mathcal N}=4)}$  
was independent of the gauge coupling, suggesting a possible special role for the curvature invariants $\tr \hat K^3 \pm h^{\mu\nu} \hat K^{\rho \sigma}W_{\mu\rho  \nu\sigma}$ in 4d theories with bulk ${\mathcal N}=4$ supersymmetry, broken in half by the boundary. Along the lines of refs.\ \cite{Drukker:2017dgn,Drukker:2017xrb,Liendo:2016ymz}, perhaps the structure of the displacement operator multiplet in ${\mathcal N}=4$ theories can shed light on this coupling independence.

Our analysis allowed for the possibility of a bulk $\theta F \wedge F$ term in the action for the photon.  We saw that this term effectively screened the interactions between the charged particles, sending $g \to g \cos(\alpha)$ where $\tan(\alpha) = g^2 \theta/ 4 \pi^2$.  Intriguingly, there is a new perturbative limit where $g \cos(\alpha)$ can be kept small even if the coupling $g$ itself is large.  As a boundary $\frac{k}{4\pi} A \wedge F$ Chern-Simons term can be absorbed by a shift in $\theta$, this perturbative large $\theta$-limit is akin to a large $k$ expansion in 3d Chern-Simons theory.

Let us conclude this paper by mentioning a number of interesting future projects that suggest themselves in light of this work, among them dynamical flavor-symmetry breaking, supersymmetric indices and localization, and extensions to 3d theories with 2d boundaries. 

When the number of electrons is even, there are suggestions \cite{Gorbar:2001qt,Kaplan:2009kr,Kotikov:2016yrn} 
that while our graphene theory is perturbatively conformal, there may be a critical value of the gauge coupling $g$ above which a mass term is spontaneously generated for the fermions which dynamically breaks some of the flavor symmetry.   Approximations for the critical $g$ come from resumming classes of Feynman diagrams.
It would be interesting to see if a similar symmetry breaking happens in supersymmetric graphene, and 
whether supersymmetry allows for a calculation of the critical $g$ with greater accuracy.%\footnote{%
%}

Given the eight supercharges and the unbroken U(1) R-symmetry of the ${\mathcal N}=2$ super graphene, it seems likely that a supersymmetric index can be constructed for this theory and also that the path integrals on special spaces, e.g.\ a hemisphere or disk with $S^3$ boundary, can be computed exactly using supersymmetric localization.  What can be learned from such an index and/or path integral?  

In 3d theories with a 2d boundary, there are again two boundary anomaly coefficients \cite{Graham:1999pm}. Let us call them $a_{(3{\rm d})}$ and $b_{(3{\rm d})}$ and write 
\be
\langle T^{\mu}{}_{\mu}\rangle^{3{\rm d}} 
= \frac{\delta(x^n)}{4 \pi} \left( a_{(3{\rm d})} \oR + b_{(3{\rm d})} \tr \hat K^2\right)\ ,
\ee  
where $\oR $ is the boundary Ricci scalar and $\tr \hat K^2= \tr K^2 - {1\over 2} K^2$.
Their relationships with the stress tensor 2-point correlation function near the boundary were discussed only recently in \cite{Herzog:2017kkj}.
We would like to find a cousin of our graphene-like theories, perhaps 3d Chern-Simons coupled to a 2d boundary fermion, in particular a cousin with an exactly marginal coupling where the boundary charge $b_{(3{\rm d})}$ can be computed perturbatively. 
(The charge $a_{(3{\rm d})}$ will not depend on the coupling in any case because it is defined from a topological invariant.)
Two groups recently have investigated dualities in 3d theories in the presence of 2d boundaries \cite{Aitken:2017nfd,Aitken:2018joi,Dimofte:2017tpi}.  
It would be interesting to use the boundary central charges to provide further evidence for dualities between field theories.  While we focused on a $U(1)$ theory here to keep the bulk non-interacting, generalizations to non-abelian gauge groups are important as well.

Clearly, there is much to be done.

\section*{Acknowledgments}

We would like to thank Nadav Drukker, 
Neil Lambert, Dario Martelli, Sameer Murthy, Martin Ro\v{c}ek, and Cristian Vergu for useful discussions.  
This work was supported in part by the U.S.\ National Science Foundation Grant PHY-1620628, 
by the U.K.\ Science \& Technology Facilities Council Grant ST/P000258/1, and by the
ERC Starting Grant N.~304806.

\newpage
\appendix

\section{Conventions}
\label{app:spinor}

We use Greek  letters $\mu$, $\nu$ for 4d bulk indices and Roman letters $A$, $B$, for 3d boundary indices.  
Our convention for the Levi-Civita tensor is that $\varepsilon^{0123}=1$.
Our metrics have mostly plus signature: $\eta^{\mu\nu} = \mbox{diag}(-,+,+,+)$ and $\eta^{AB} = \mbox{diag}(-,+,+)$.  
Our gamma matrices satisfy the Clifford algebras:
\be
\{\gamma^\mu, \gamma^\nu \} = - 2 \eta^{\mu\nu} \; \; \; \mbox{and} \; \; \;
\{\Gamma^A, \Gamma^B \} = - 2 \eta^{AB} \ .
\ee
In 4d, the ``fifth'' gamma matrix is defined to be $\gamma^5 \equiv \gamma^0\gamma^1\gamma^2\gamma^3$.
Two gamma matrix identities useful for verifying SUSY of the 4d vector mulitplet are
\be
\label{gammaone}
\frac{1}{2} [\gamma_\mu, \gamma_\nu] \gamma_\rho &=& \eta_{\mu\rho} \gamma_\nu - \eta_{\nu \rho} \gamma_\mu  +  \varepsilon_{\mu\nu\rho \sigma} \gamma^\sigma \gamma^5 \ , \\
\label{gammatwo}
\frac{1}{2} \gamma_\rho [\gamma_\mu, \gamma_\nu] &=& -\eta_{\mu\rho} \gamma_\nu + \eta_{\nu \rho} \gamma_\mu + \varepsilon_{\mu\nu\rho \sigma} \gamma^\sigma \gamma^5 \ .
\ee

 For definiteness, let us choose a basis for the 4d gamma matrices:
\be
 \gamma^\mu = \left(\begin{array}{cc}
 0 & \sigma^\mu \\ \bar{\sigma}^\mu & 0
 \end{array} \right)\ ,
 \; \; \;
 \gamma^5 
  = \left(\begin{array}{cc} -i & 0 \\ 0 & i
 \end{array} \right)
 \ ,
\ee
where $\sigma^\mu = (-1, \vec \sigma)$ and $\bar \sigma^\mu  = (-1, - \vec \sigma)$.  
Note that $(\gamma^0)^\dagger = \gamma^0$ while $(\gamma^i)^\dagger = - \gamma^i$.  Defining the cospinors in the usual way $\bar \psi = \psi^\dagger \gamma^0$, from these identities and the anti-commutation relations one can deduce $\overline{\gamma^\mu \psi} = \bar \psi \gamma^\mu$.  
Furthermore $(\gamma^0)^T = \gamma^0$ and $(\gamma^2)^T = \gamma^2$ while $(\gamma^1)^T =  -\gamma^1$ and $(\gamma^3)^T = - \gamma^3$.

The photino is a Majorana spinor and thus satisfies the reality condition $\lambda = \lambda^C \equiv C \bar \lambda^T$ (or equivalently $\bar \lambda = \lambda^T C$) where $C \equiv i \gamma^0 \gamma^2 = -C^T = -C^\dagger = C^* = -C^{-1} = \mbox{diag}(i \sigma^2, -i \sigma^2)$.  
The reality constraint implies a handful of bilinear identities useful for demonstrating SUSY:
\be
\label{Majbil}
\bar s_1 M s_2 &=& s_1^T C M s_2
= -s_2^T C C^{-1} M^T C^T s_1
=  \bar s_2 C^{-1} M^T C s_1 \ .
\ee
Useful special cases are 
\be \label{MT_conj}
C^{-1} M^T C = \begin{cases}
M &   M = 1 , \; \gamma_5 \gamma_\mu \ , \; \gamma_5 \ , \\
-M   &  M = \gamma_\mu \ , \;  [\gamma_\mu, \gamma_\nu]  \ , \;  \gamma_5 [\gamma_\mu, \gamma_\nu] \ .
\end{cases}
\ee
Another useful relation is  $C \gamma_\rho [\gamma_\mu, \gamma_\nu] C^{-1} = ([\gamma_\mu, \gamma_\nu] \gamma_\rho)^T$.

\subsubsection*{Reducing 4d Majorana Spinors to 3d}

The relation between the 4d and 3d spinor expressions can be obtained as follows. Given the 3d gamma matrices $\Gamma^A$, a 4d Clifford algebra can be constructed as
\be \label{3d_gamma_emb}
\t \gamma^A = \begin{pmatrix} \Gamma^A &0 \\ 0 & - \Gamma^A \end{pmatrix} \ ,
\qquad
\t \gamma^n = \begin{pmatrix} 0 & e^{i\eta} \\ -e^{-i\eta} & 0 \end{pmatrix} \ .
\ee
In this basis the projectors $\Pi_\pm = \frac{1}{2}(1\pm \beta)$ with $\beta = i\gamma^n \gamma^5 e^{\eta\gamma^5}$ are diagonal and commute with the tangential 4d gamma matrices. The relation $\Pi_\pm e^{-\eta\gamma^5} = e^{-\eta\gamma^5}\overline \Pi_\pm$ suggests the identification $\t \gamma^\mu=e^{-\eta\gamma^5} \gamma^\mu$. The transformation that diagonalizes the projectors is given by 
\be \label{}
U 
= 
\frac{1}{\sqrt2} 
\begin{pmatrix}
1 &0 \\
0 & \sigma^n
\end{pmatrix}
\left( 1- i \beta\gamma^5 \right) 
= 
\frac{1}{\sqrt2} \begin{pmatrix} 1 & -e^{i\eta} \sigma^n \\ e^{-i\eta}  & \sigma^n \end{pmatrix} \ ,
\ee
written explicitly in the Weyl basis. 
A 3d spinor $\psi$ is identified, in the basis where the projectors are diagonal, with $(\psi \,\,0 )^T$ when it is embedded as an eigenvalue 1 of $\Pi_+$. In the Weyl basis it takes the form
\begin{align} \label{psiembedding}
e^{\frac{i\eta}{2}}U^{-1} \begin{pmatrix} \psi \\  0\end{pmatrix}
=
\frac{1}{\sqrt2}e^{- \frac{\eta}{2}\gamma^5} \begin{pmatrix} \psi \\  \bar\sigma^n \psi \end{pmatrix} \ ,
\end{align}
with the extra phase in $U$ chosen for convenience. The actual form of the embedded 3d spinor in the Weyl basis is never needed and we can trade it for $\psi$ without confusion.

Let us develop the dictionary for converting 4d expressions to 3d ones. It follows from the relations above that we can exchange
\be
\Pi_\pm \t\gamma^{A}  =  \pm \Gamma^A \ ,
\qquad
\Pi_\pm \t\gamma^n  =  \pm i \gamma^5 \Pi_\mp \ .
\ee
The 3d bar is related to the 4d bar by
\be \label{app_3d_bar}
\bar\lambda 
\equiv 
\lambda^\dagger \gamma^0 
= 
\lambda^\dagger \t \gamma^0 e^{-\eta\gamma^5} 
\equiv 
\t\lambda e^{-\eta\gamma^5} \ ,
\ee
and the 3d bar has the property $\t{\Pi_\pm \lambda} = \t\lambda \Pi_\pm$. The 3d charge conjugation matrix can be identified 
such that $\t\lambda = \lambda^T \t C$. Comparing with $\bar\lambda = \lambda^T C$ and \eqref{app_3d_bar} leads to
\be
\t C = C e^{\eta\gamma^5} \ , 
\ee
with the properties
\be
\t C^T = - \t C \ , 
\qquad
\t C^\dagger = \t C^{-1} =  e^{-\eta\gamma^5} C^{-1} \neq - \t C \ ,
\qquad
\t C \,\t \gamma^\mu \t C^{-1} = - \t \gamma^{\mu \, T} \ .
\ee

In verifying SUSY on the boundary, a Fierz rearrangement identity is required,
\be
(\t \lambda \psi) (\t \psi \chi)-(\t \chi \psi) (\t \psi \lambda) =   (\t \lambda \Gamma_A \chi) (\t \psi \Gamma^A \psi) \ .
\ee

\subsection*{Symplectic Majorana Fermions}

In the case of $\mathcal{N}=2$ supersymmetry the number of supercharges is doubled. We now have $Q^i$ where $i=1,2$ is a fundamental $SU(2)$ index. It is lowered and raised according to the conventions $Q_j = \varepsilon_{ji} Q^i$ and $Q^i = \varepsilon^{ij} Q_j$ with $\varepsilon^{12} = -\varepsilon_{12} =1$. We also use the convention $(Q^i)^* = Q_i$ so that upper and lower index contractions form invariants. The generators of the R-symmetry are the Pauli matrices denoted by $(\vec \tau)^i{}_j$ and satisfy the relation $\varepsilon_{ik} (\vec\tau)^k{}_j  = \varepsilon_{jk}(\vec\tau)^k{}_i$. For $\mathcal{N}=2$, the previously used Majorana condition is incompatible with the $SU(2)$ symmetry. 
Instead we introduce the symplectic Majorana condition, defining a new charge conjugation matrix by $C_+ = i\gamma^5 C$, where the plus serves to indicate the relation $C_+ \gamma^\mu C_+^{-1} = (\gamma^\mu)^T$ contrary to \eqref{MT_conj}. 
\be
\bar \lambda_i =  \varepsilon_{ij} \lambda^{j T} C_+ \ . 
\ee
There are a host of bilinear relations necessary for demonstrating SUSY.  The analogs of 
the Majorana relation (\ref{Majbil}) are  
\be
\bar s_{1i} M {s_2}^i &=& - \bar s_{2i} C_+^{-1} M^T C_+ {s_1}^i \ , \\
\bar s_{1i} M {\tau^i}_j{s_2}^j &=& \bar s_{2i} C_+^{-1} M^T C_+ {\tau^i}_j{s_1}^j \ .
\ee
Rewriting $C_+^{-1} M^T C_+ = i\gamma^5 C^{-1} M^T C i \gamma^5$, using the results in \eqref{MT_conj}, one obtains
\be
\bar \epsilon_i \lambda^i &=& - \bar \lambda_i \epsilon^i \ , \\
\bar \epsilon_i \gamma^\mu \lambda^i &=& - \bar \lambda_i \gamma^\mu \epsilon^i \ , \\
\bar \epsilon_i \gamma^5 \lambda^i &=& - \bar \lambda_i \gamma^5 \epsilon^i \ , \\
\bar \epsilon_i \gamma^5 \gamma^\mu \lambda^i &=& \bar \lambda_i \gamma^5 \gamma^\mu \epsilon^i \ , \\
\bar \epsilon_i \gamma^{\mu\nu} \lambda^i &=& \bar \lambda_i \gamma^{\mu\nu} \epsilon^i \ , \\
\bar \epsilon_i \gamma^5 \gamma^{\mu\nu} \lambda^i &=& \bar \lambda_i \gamma^5 \gamma^{\mu\nu} \epsilon^i \ .
\ee
The second relation introduces an additional minus sign.

Similar to the ${\mathcal N}=1$ case, we can write down the eigenvectors of the projection matrices $\Pi_\pm$ in a basis where $\gamma^5$ is diagonal, although for the most part we do not need them.  For simplicity, focus on the case where $\vec v = (0,0,1)$ and the action of the SU(2) generators $\vec \tau$ is already diagonalized.  In this case, 
\be
\Pi_+ \epsilon^1 = \epsilon^1 &=& \frac{1}{\sqrt{2}} e^{- \frac{\eta}{2}\gamma^5} \begin{pmatrix} \epsilon \\  \bar\sigma^n \epsilon \end{pmatrix} \ , 
\\
\Pi_+ \epsilon^2 = \epsilon^2 &=&- \frac{1}{\sqrt{2}} e^{- \frac{\eta}{2}\gamma^5} \begin{pmatrix} \epsilon^c \\  -\bar\sigma^n \epsilon^c \end{pmatrix} \ , ~~~
\epsilon^c = - \t C_+^{-1} \t \epsilon^{\, \, T} \ .
\ee

\section{Feynman Rules 
}
\label{FR}

The Feynman rules can be read from the ${\mathcal N}=1$ and ${\mathcal N}=2$ Lagrangians.\footnote{%
Note the $\lambda \psi \phi$ vertices are written for the ${\mathcal N}=1$ theory.  In the ${\mathcal N}=2$ theory, they pick up a factor of $\sqrt{2}$ and $\lambda$ is no longer Majorana.  
}
We specialize to propagators where at least one of the two points is on the boundary.  
The rules are as follows: 
\\

\noindent
{\it Propagators:}
\\

\begin{tabular}{cclcccl}
\multicolumn{7}{c}{
$A_\mu$ (Feynman gauge):  
\begin{tikzpicture}
\draw[decorate ,decoration={coil,aspect=0}] (0,0) -- (3,0); 
\label{Ap}
\end{tikzpicture}
~~
$=-i \frac{e^{-p y}}{p} \eta^{AB}$ 
}
\\
\\
$\lambda$: 
&
\begin{tikzpicture}
\draw (0,-2.5) -- (3,-2.5);
\draw [decoration={markings,mark=at position 1 with {\arrow[scale=2.3,>=stealth]{>}}},postaction={decorate}] (1.8,-2.5) -- (2,-2.5);
\draw [decoration={markings,mark=at position 1 with {\arrow[scale=2.3,>=stealth]{<}}},postaction={decorate}] (1.2,-2.5) -- (1.4,-2.5);
\label{lp}
\end{tikzpicture}
&
$=\Pi_+ \frac{i \gamma^A p_A e^{-p y}}{p}$ 
&
~~~
&
$X$~and~$Y$: 
&
\begin{tikzpicture}
\draw[dashed] [thick]
(0,-2.5) -- (3,-2.5);
\label{xp}
\end{tikzpicture}
&
$= \frac{-i e^{-py}}{p}$
\\
\\
$\psi$: &
\begin{tikzpicture}
\draw
(0,-2.5) -- (3,-2.5);
\draw [decoration={markings,mark=at position 1 with {\arrow[scale=2.4,>=stealth]{>}}},postaction={decorate}] (1.57,-2.5) -- (1.7,-2.5);
\label{psip}
\end{tikzpicture}
&
$=\frac{i \Gamma^A p_A}{p^2}$
&
~~~
&
$\phi$:
&
\begin{tikzpicture}
\draw[dotted][thick]
(0,-2.5) -- (3,-2.5);
\draw [decoration={markings,mark=at position 1 with {\arrow[scale=2.4,>=stealth]{>}}},postaction={decorate}] (1.57,-2.5) -- (1.7,-2.5);
\label{phip}
\end{tikzpicture}
&= $ \frac{-i}{p^2}$
\end{tabular}

%\pagebreak
\vskip 0.2in

\noindent
{\it Vertices:}\\

\begin{tabular}{lcl}

%~~~
\begin{tikzpicture}
\draw (4,-2.5) node[above] {$= i g \Gamma^B$};
\draw[decorate ,decoration={coil,aspect=0}] (1.5,-1) -- (1.5,-2.6); 
\draw
(0,-2.6) -- (3,-2.6);
\draw [fill] (1.5,-2.6) circle [radius=.05];
\draw [decoration={markings,mark=at position 1 with {\arrow[scale=2.4,>=stealth]{>}}},postaction={decorate}] (0.75,-2.6) -- (0.85,-2.6);
\draw [decoration={markings,mark=at position 1 with {\arrow[scale=2.4,>=stealth]{>}}},postaction={decorate}] (2.25,-2.6) -- (2.35,-2.6);
\label{v1}
\end{tikzpicture}
&
~~~~~
&
\begin{tikzpicture}
\draw (4.5,-2.5) node[above] {$= -i g (p+p')_A$};
\draw[decorate ,decoration={coil,aspect=0}] (1.5,-1) -- (1.5,-2.6); 
\draw[dotted][thick] 
(0,-2.6) -- (3,-2.6);
\draw [decoration={markings,mark=at position 1 with {\arrow[scale=2.4,>=stealth]{>}}},postaction={decorate}] (0.75,-2.6) -- (0.85,-2.6);
\draw [decoration={markings,mark=at position 1 with {\arrow[scale=2.4,>=stealth]{>}}},postaction={decorate}] (2.25,-2.6) -- (2.35,-2.6);
\draw[thin]
(0.3,-2.2) -- (1.2,-2.2);
\draw[thin]
(1.8,-2.2) -- (2.7,-2.2);
\draw [decoration={markings,mark=at position 1 with {\arrow[scale=1.4,>=stealth]{>}}},postaction={decorate}] (1.1,-2.2) -- (1.2,-2.2);
\draw [decoration={markings,mark=at position 1 with {\arrow[scale=1.4,>=stealth]{>}}},postaction={decorate}] (2.6,-2.2) -- (2.7,-2.2);
\draw (0.75,-2.1) node[above] {$p'$};
\draw (2.25,-2.1) node[above] {$p$}; 
\draw [fill] (1.5,-2.6) circle [radius=.05];
\end{tikzpicture}
\label{v2}
%\end{figure}
\\
\\
\begin{tikzpicture}
\draw (4,-1) node[above] {$= -2 i g^2 \eta_{AB}$};
\draw[decorate ,decoration={coil,aspect=0}] (-0.5,0.3) -- (1,-0.5); 
\draw[decorate ,decoration={coil,aspect=0}] (-0.5,-1.3) -- (1,-0.5); 
\draw[dotted][thick] (1,-0.5) -- (2.4,0.3);
\draw[dotted][thick] (1,-0.5) -- (2.4,-1.3);
\draw [fill] (1,-0.5) circle [radius=.05];
\draw [decoration={markings,mark=at position 1 with {\arrow[scale=2.4,>=stealth]{>}}},postaction={decorate}] (1.7,-0.09) -- (1.8,-0.02);
\draw [decoration={markings,mark=at position 1 with {\arrow[scale=2.4,>=stealth]{>}}},postaction={decorate}] (1.7,-0.89) -- (1.6,-.82);
\end{tikzpicture}
\label{v3}
&
~~~~~
&
\begin{tikzpicture}
\draw (3.5,-2.3) node[above] {$= ig$};
\draw[dashed][thick] (1.5,-1) -- (1.5,-2.6); 
\draw 
(0,-2.6) -- (3,-2.6);
\draw [fill] (1.5,-2.6) circle [radius=.05];
\draw [decoration={markings,mark=at position 1 with {\arrow[scale=2.4,>=stealth]{>}}},postaction={decorate}] (0.75,-2.6) -- (0.85,-2.6);
\draw [decoration={markings,mark=at position 1 with {\arrow[scale=2.4,>=stealth]{>}}},postaction={decorate}] (2.25,-2.6) -- (2.35,-2.6);
\end{tikzpicture}
\label{v5}
\\
\\
\begin{tikzpicture}
\draw (3.7,-2.4) node[above] {$=g$};
\draw [decoration={markings,mark=at position 1 with {\arrow[scale=2.2,>=stealth]{>}}},postaction={decorate}] (1.5,-2.1) -- (1.5,-2.2);
\draw [decoration={markings,mark=at position 1 with {\arrow[scale=2.2,>=stealth]{<}}},postaction={decorate}] (1.5,-1.6) -- (1.5,-1.7);
\draw(1.5,-1) -- (1.5,-2.6); 
\draw[dotted][thick] 
(0,-2.6) -- (1.5,-2.6);
\draw(1.5,-2.6) -- (3,-2.6);
\draw [fill] (1.5,-2.6) circle [radius=.05];
\draw [decoration={markings,mark=at position 1 with {\arrow[scale=2.4,>=stealth]{>}}},postaction={decorate}] (0.75,-2.6) -- (0.85,-2.6);
\draw [decoration={markings,mark=at position 1 with {\arrow[scale=2.4,>=stealth]{>}}},postaction={decorate}] (2.25,-2.6) -- (2.35,-2.6);
\end{tikzpicture}
\label{v4}
&
~~~~~
&
\begin{tikzpicture}
\draw (3.5,-2.4) node[above] {$=-g$};
\draw [decoration={markings,mark=at position 1 with {\arrow[scale=2.2,>=stealth]{>}}},postaction={decorate}] (1.5,-2.1) -- (1.5,-2.2);
\draw [decoration={markings,mark=at position 1 with {\arrow[scale=2.2,>=stealth]{<}}},postaction={decorate}] (1.5,-1.6) -- (1.5,-1.7);
\draw(1.5,-1) -- (1.5,-2.6); 
\draw[dotted][thick] 
(0,-2.6) -- (1.5,-2.6);
\draw(1.5,-2.6) -- (3,-2.6);
\draw [fill] (1.5,-2.6) circle [radius=.05];
\draw [decoration={markings,mark=at position 1 with {\arrow[scale=2.4,>=stealth]{>}}},postaction={decorate}] (0.75,-2.6) -- (0.65,-2.6);
\draw [decoration={markings,mark=at position 1 with {\arrow[scale=2.4,>=stealth]{>}}},postaction={decorate}] (2.25,-2.6) -- (2.15,-2.6);
\end{tikzpicture}
\label{v4b}
\\
\\

\begin{tikzpicture}
\draw (3.5,-1.1) node[above] {$= -2ig^2$};
\draw[dashed][thick] (-0.5,0.3) -- (1,-0.5); 
\draw[dashed][thick] (-0.5,-1.3) -- (1,-0.5); 
\draw[dotted][thick] (1,-0.5) -- (2.4,0.3);
\draw[dotted][thick] (1,-0.5) -- (2.4,-1.3);
\draw [fill] (1,-0.5) circle [radius=.05];
\draw [decoration={markings,mark=at position 1 with {\arrow[scale=2.4,>=stealth]{>}}},postaction={decorate}] (1.7,-0.09) -- (1.8,-0.02);
\draw [decoration={markings,mark=at position 1 with {\arrow[scale=2.4,>=stealth]{>}}},postaction={decorate}] (1.7,-0.89) -- (1.6,-.82);
\end{tikzpicture}
\label{v6}
&
~~~~~
&
\begin{tikzpicture}
\draw (3.7,-2.4) node[above] {$=-ipg$};
\draw[dashed][thick] (1.5,-1) -- (1.5,-2.6); 
\draw [dotted][thick]
(0,-2.6) -- (3,-2.6);
\draw [fill] (1.5,-2.6) circle [radius=.05];
\draw [decoration={markings,mark=at position 1 with {\arrow[scale=2.4,>=stealth]{>}}},postaction={decorate}] (0.75,-2.6) -- (0.85,-2.6);
\draw [decoration={markings,mark=at position 1 with {\arrow[scale=2.4,>=stealth]{>}}},postaction={decorate}] (2.25,-2.6) -- (2.35,-2.6);
\end{tikzpicture}
\label{v7}
\end{tabular}

\end{document}